%% file: main.tex
\documentclass[a4paper,12pt]{article}
\usepackage{multirow}
\usepackage{amsmath}
\usepackage{arxiv}
\usepackage{float}
\usepackage[utf8]{inputenc} 
\usepackage[T1]{fontenc}    
\usepackage{natbib}
\usepackage[colorlinks=true,allcolors=blue]{hyperref}    
\usepackage{url}  
\usepackage{array}
\usepackage{booktabs}       
\usepackage{amsfonts}       
\usepackage{nicefrac}       
\usepackage{microtype} 
\usepackage{numprint}
\usepackage{lipsum}
\usepackage{rotating}
\usepackage[nodayofweek]{datetime}
\usepackage{xcolor}
\usepackage{color}
\usepackage{titlesec}
\usepackage{appendix}
\titleformat*{\section}{\LARGE\bfseries}
\titleformat*{\subsection}{\Large\bfseries}
\titleformat*{\subsubsection}{\large\bfseries}
\titleformat*{\paragraph}{\large\bfseries}
\titleformat*{\subparagraph}{\large\bfseries}
\usepackage{chngcntr}
\usepackage{listings}
\usepackage[flushleft]{threeparttable}
\usepackage[ruled,vlined]{algorithm2e}
\usepackage{longtable}
\usepackage{graphicx}
\AtBeginDocument{\hypersetup{pdfborder={0 0 0}}}
\definecolor{mygreen}{rgb}{0,0.6,0}
\definecolor{mygray}{rgb}{0.5,0.5,0.5}
\definecolor{mymauve}{rgb}{0.58,0,0.82}
\usepackage[bf]{caption}
\newdateformat{monthyeardate}{%
  \monthname[\THEMONTH], \THEYEAR}
  \pdfoutput=1  

\title{Asset Prices and Capital Share Risks: Theory and Evidence}
\author{
\large{Joseph P. Byrne}\thanks{Edinburgh Business School, School of Social Sciences. Heriot-Watt University. Email: J.P.Byrne@hw.ac.uk.} \\\\\\\\\\
   \And
\large{Boulis M. Ibrahim} 
\thanks{Edinburgh Business School, School of Social Sciences.
Heriot-Watt University. Email: B.M.Ibrahim@hw.ac.uk.}\\\\\\\\\\
   \And
     \large{Xiaoyu Zong}\thanks{Correspondence author: Xiaoyu Zong. Edinburgh Business School, School of Social Sciences. Heriot-Watt University. Email: xz68@hw.ac.uk.} \\\\\\\\}

\makeatletter
\let\TPT@hookin\@gobble
\let\TPT@hookarg\@gobble
\makeatother

\begin{document}
\linespread{1.25}
\maketitle

\large
\begin{abstract}
An asset pricing model using long-run capital share growth risk has recently been found to successfully explain U.S. stock returns. Our paper adopts a recursive preference utility framework to derive an heterogeneous asset pricing model with capital share risks.While modeling capital share risks, we account for the elevated consumption volatility of high income stockholders. Capital risks have strong volatility effects in our recursive asset pricing model. Empirical evidence is presented in which capital share growth is also a source of risk for stock return volatility. We uncover contrasting unconditional and conditional asset pricing evidence for capital share risks.
\end{abstract}

\begin{center}
        \begin{minipage}{.85\linewidth}{
\smallbreak\keywords{Asset Pricing, Capital Share,  Recursive Preference, Consumption Growth, Bayesian Methods.}
\JELs{C21, C30, E25, G11, G12.}}\end{minipage}
\end{center}
\setcounter{footnote}{0}
\newpage
\input{chapters/Introduction.tex}
\input{chapters/lit_review.tex}

\input{chapters/Methodology/Methodology.tex}

\input{chapters/4-Data-Priors/4-Data-Prior.tex}

\input{chapters/Results/Results.tex}

\input{chapters/further_discussion.tex}

\input{chapters/conclusion.tex}

\newpage
\bibliographystyle{apalike}  
\bibliography{main.bbl}  

\newpage
\clearpage
\pagenumbering{arabic}
\appendix
\numberwithin{equation}{section}
\input{chapters/appendix/appendix.tex}
\end{document}

%% file: chapters/Introduction.tex
\section{Introduction}
Leading asset pricing theories frequently assume a single representative agent when seeking to model expected returns. For instance, Breeden's (\citeyear{breeden1979intertemporal}) consumption based approach adopts a representative agent, allowing aggregate consumption growth to systematically price returns. However, stock returns are considerably more volatile than aggregate consumption growth. This empirical observation is a cornerstone of the equity premium puzzle, see \cite{mehra1985equity} and \cite{breeden2014consumption}. When endeavoring to explain the failure of consumption-based asset pricing models, one can relax the homogeneous agent assumption (see \cite{constantinides1996asset}, \cite{chabi2014aggregation} and \cite{lettau2019capital}). Observed differences in stock market participation, resources and/or preferences justify heterogeneous agent asset pricing models. Indeed, \cite{campbell1993trading} emphasize a focus upon heterogeneous agent asset price model, since it is shareholder consumption that matter for stock returns.

Important recent work by \cite{lettau2019capital} adopts heterogeneous agents and proposes a capital share risk factor that maps shocks to the assets of high income stock holders' assets and consumption to explain U.S. returns. Capital risks account for limited stock market participation and proxy the concentration of wealth. \cite{lettau2019capital} present evidence that a capital risk factor explains expected returns, and empirically dominate aggregate consumption growth and the Fama and French (\citeyear{fama1993common}) factors. In developing a general equilibrium model with limited stock market participation and inequality, \cite{toda2018equity} highlight that rising wealth holdings of the richest one percent predict excess stock returns. Their asset pricing model allows for heterogeneity in risk aversion or beliefs. See also theoretical asset pricing models accounting for heterogeneous beliefs under recursive preferences by \cite{borovivcka2020survival}.       

While explaining excess equity returns has been the focus of much academic research, elevated stock return variability has also been considered by homogeneous agent models. Seeking to resolve asset pricing puzzles associated with standard models, \cite{bansal2004risks} develop a consumption representative agent model with long-run risks based upon the recursive preference utility framework of \cite{epstein1989substitution}.  The asset pricing puzzles highlighted by \cite{bansal2004risks} include high conditional volatility of market returns and a negative risk premium on consumption volatility. 

In this paper we develop a heterogeneous agent stock market model and test its predictions. Going beyond the homogeneous framework developed by \cite{bansal2004risks}, our theoretical work generalizes the heterogeneous agent assumption of \cite{lettau2019capital} and allows us to reconsider their empirical evidence, accounting in particular for time-varying risk prices and equity return volatility.Our asset pricing model has multiple economic agents: in particular, high and low income stockholders with different consumption patterns. Using United States wealth distribution data from \cite{saez2016wealth}, \cite{lettau2019capital} identify that high and low income stockholders' consumption behaviour responds differently to capital share growth. Given that high income stockholders consume primarily out of their wealth, capital share growth can also explain the elevated consumption variability of the richest cohorts. Therefore, our theory captures heterogeneity in the volatility of stockholder consumption growth, and this drives the relationship between capital share growth and equity returns. The impact of consumption volatility of wealthy stockholders on the whole market is captured by capital share change.

One novelty of our model is that the same risk factor is analysed separately using conditional and unconditional expectations. Inspired by \cite{campbell2000explaining}, the capital share risk factor in our model is linked to stock returns under both unconditional and conditional expectations. The contrasting impact of the capital share factor on equity returns under conditional and unconditional settings also serves as a potential explanation of the weakness of the consumption-based CAPM under conditional estimation \citep{campbell2000explaining}. Capital share growth is priced only under unconditional expectations, as found by \cite{lettau2019capital}, and capital share variability is proposed in our paper to capture long-run market volatility. Therefore, our framework posits that capital risks enter the equity return variance equation under conditional estimation, while they enter the mean equation under unconditional estimation. 

Motivated by our theoretical framework in which capital shares impact not only the mean but also the variance of return dynamics, we test the pricing power of capital risks in a more general setting than that of \cite{lettau2019capital}. To avoid firm effects, we first use bootstrapped cross-sectional regressions to estimate the level and volatility effects of capital share risks in asset prices. The capital share risk price is adopted as a benchmark for pricing power in this case. To then test the conditional equity premium dynamics, we investigate the conditional capital share risk prices using the rolling-window Fama-MacBeth procedure used by \cite{lewellen2006conditional} and the Bayesian asset price estimation from \cite{bianchi2017macroeconomic}. Additionally, our paper estimates the impact of capital share growth on return volatility using rolling-window multiplicative GARCH. Finally, we test the capital risk variability as the long-run risk factor in the mean equation for U.S. stock returns using a bootstrap procedure, as we do for testing the capital share factor. This alternative capital variability factor empirically dominates the standard capital share growth factor of \cite{lettau2019capital}.

The paper is structured as follows: Section 2 reviews standard asset pricing models and the capital share risk factor proposed by \cite{lettau2019capital}. Section 3 presents a theoretical asset pricing model with recursive preferences and heterogeneous agents, in which consumption volatility operates through capital share risk factors. Section 4 sets out empirical methodologies for estimating overall pricing power, and the level and volatility effects of the capital share factor. Section 5 presents the data and Section 6 presents evidence on the impact of capital share risks. Section 7 proposes and tests the capital share variability factor. Finally, section 8 concludes on the capital share variability and the empirical evidence.

%% file: chapters/lit_review.tex
\section{Asset Pricing Theory and Capital Share Growth}
\label{section:lit}
\noindent
Modern asset pricing models describe the relationship between risk exposure and expected returns. Expected returns equal the sum of the risk free rate and the excess returns of associated risk factors. In capital asset pricing models, the stochastic discount factor (SDF) links the present value and the future cash flow of an asset, and the price of an asset can be computed by the expectation of total future cash flows discounted by that discount factor. 

An asset pricing model can be seen as a special case of the following relationship:
\begin{equation}
    E_t(M_{t+1}r_{t+1})=0
    \label{eq:sdf}
\end{equation}
where $M_{t+1}$ denotes the discount factor, $r_{t+1}$ denotes asset excess returns and $E_t$ is the conditional expectation given information at time $t$. The form of the SDF relies heavily on the assumptions made by different CAPMs \citep{cochrane2009asset}. Equation (\ref{eq:sdf}) is operationalized by agents' expectation formation and their utility or preferences function(s).

The consumption‐based capital asset pricing model (CCAPM) developed by \cite{breeden1979intertemporal} states that, with a representative agent assumption, the SDF is based on the marginal rate of substitution over aggregate household consumption. The CCAPM assumes the SDF is equal to the time-discount factor ($\delta$) multiplied by the ratio of the marginal utility of aggregate consumption tomorrow $U'(C_{t+1})$ and today $U'(C_{t})$ as shown in equation (\ref{eq:ccapm_sdf}), where $U$ denotes the utility function of the representative agent, and $C_t$ denotes consumption at time $t$.
\begin{equation}
    M_{t+1}=\delta\frac{U'(C_{t+1})}{U'(C_t)}
    \label{eq:ccapm_sdf}
\end{equation}

With an homogeneous agent and power utility, expected returns can be priced by aggregate consumption growth. Agents are however more reasonably considered to be heterogeneous, due to imperfect
risk-sharing, concentrated wealth and limited stock holder participation. And the CCAPM may not
perform well empirically, see \cite{breeden2014consumption}. According to \cite{lettau2019capital}, for the wealthiest households, relative to the least wealthy,  aggregate consumption
volatility multiplied by income share is considerably high. 
Stock market wealth is highly concentrated, since the top 5\% of the wealth distribution owns over 70\% of stocks. The least wealthy typically own no equity, and their consumption comes almost entirely from labor income. Therefore, aggregate consumption growth also fails to capture redistribution risk.\footnote{Wealth re-distributions between stockholders and workers.}
Capital share growth better reflects the consumption of stockholders, while accounting for stockholder heterogeneity and redistributive shocks (see \cite{greenwald2014origins} and \cite{lettau2019capital}). The income shares, and therefore consumption patterns, of wealthy capital owners is well represented by the capital share.

\cite{lettau2019capital} then proposes a linear approximation of the asset price SDF using the capital share factor:
\begin{equation}
    M_{t+1}=a+b_{1}(\frac{C_{t+1}}{C_{t}}-1)+b_{2}(\frac{KS_{t+1}}{KS_{t}}-1)+\mu_{t+1}
    \label{eq:ks_apx}
\end{equation}
Equation (\ref{eq:ks_apx}) describes the richest household case in a stylised economy where workers are assumed to be absent from risky asset markets. $KS_t$ denotes capital share, $a$ is related to the time-discount factor (i.e. $\alpha = 1 + ln(\delta)$), while model parameters $b_{1}$ and $b_{2}$ are related to the risk aversion of consumers. Evidence presented by \cite{lettau2019capital} indicates that capital share growth explains U.S. asset prices and restricting aggregate consumption growth to have no effect (i.e. $b_1=0$) is a reasonable assumption. Approximation error is denoted by $\mu$ and explains other factors. 

The model proposed by \cite{lettau2019capital} does not allow for time variation of the capital share parameters. Empirically risk factor loadings however may vary over time, and conditional asset pricing models can be justified theoretically (see \cite{jensen1968performance}, \cite{jagannathan1996conditional} and \cite{lewellen2006conditional}). For instance, the static CCAPM fails to capture the effect of time-varying investment opportunities \citep{lettau2001resurrecting}. The non-zero unconditional price anomalies do not necessarily indicate non-zero conditional alphas, given time-varying factor loadings that are correlated with the equity premium or market volatility \citep{lewellen2006conditional}. Our paper generalises the role of capital share by testing both unconditional and conditional approaches.

\input{chapters/Results/proof}

%% file: chapters/Results/proof.tex
 \section{A Capital Share Asset Pricing Model}
 \label{section:theory}
 \subsection{The Recursive Preference Utility Framework}
Capital share can be motivated as an asset pricing risk factor in a stylized economy with heterogeneous agents, see \cite{lettau2019capital}.  In this economy, capital investors own the entire corporate sector, while workers do not participant in the stock market. Capital share is assumed to play a role through its influence on stockholders' consumption. Therefore, in this case, the stochastic discount factor is represented by the utility function of investors of the top wealth distribution. 
 
Our capital share asset pricing model relaxes the Lettau et al.'s (\citeyear{lettau2019capital}) assumption that capital share only impacts the top wealth distribution of stockholders to derive a more general case. Here capital share growth is assumed to influence both the high and the low income stockholder groups, since the stock market and the wealth weighted participation rates are not identical (see \cite{lettau2019capital}).  Our model assumes three income groups, which are high income stockholders, low income stockholders and labour workers. Labour workers are assumed to be absent from the stock market as in \cite{lettau2019capital}. We assume for simplicity a constant relative share of high versus low income stockholders.\footnote{Individual can move between income groups, but the population sizes to move into and out from each income group are the same.} The population weight of the high income stockholders is denoted by $w^H$ and that of the low income stockholders is denoted by $w^L$. We then introduce these high and low income stockholders into a recursive asset pricing model.

Over an infinite horizon, the CCAPM in equation (\ref{eq:ccapm_sdf}) is nested as a special case in asset return models derived from a recursive preference utility framework proposed by \cite{epstein1989substitution}. This recursive framework permits risk attitudes to be disentangled from the degree of intertemporal substitutability, and addresses the importance of consumption uncertainty in asset pricing \citep{epstein1989substitution}. Epstein and Zin (\citeyear{epstein1989substitution}) assume agents are homogeneous in the market, and our capital share model relaxes this assumption. In our model, the homogeneous assumptions for the recursive framework are satisfied within each stockholder group.

\subsection{Innovations to the Pricing Kernel}
Going beyond the homogeneous agent model of \cite{bansal2004risks}, the stockholder consumption growth $G_{t+1}$ is underpinned by the particular consumption patterns of different investors.\footnote{In our paper, $G^n_t$ denotes the consumption growth of agent $n$ calculated from $G^n_t=C^n_{t+1}/C^n_t$, where $C^n_t$ is the consumption at time $t$ of agent $n$. } According to \cite{lettau2019capital}, the consumption growth rate of the high income stockholders is more volatile than that of those who derive income from wages. Intuitively, the top of the wealth distribution has a larger discretionary consumption on luxury goods 
linked to volatile asset prices, while workers spend a larger proportion on the same essential goods each month. The capital share growth rate is strongly and positively correlated with the consumption growth rate of the high income group, while strongly negative correlated with that of the low income group \citep{lettau2019capital}. Accordingly, we define $\bar{G_t}$ as the weighted average of consumption growth of labour workers and the growth of the high and low income stockholders: 
\begin{equation}
    \bar{G_t}\approx r^p (w^HG_t^H+w^LG_t^L)+(1-r^p)G_t^{W}
    \label{EQ:G_OVERALL}
\end{equation}
where $r^p$ is the stock market participation rate, and $G_t^{W}$ is the consumption of labour workers without stocks. The aggregate consumption growth $\bar{G}$ is independent from how the income groups are defined. Our heterogeneous model focuses upon the high ($G^H_t$) and low ($G^L_t$) income stockholder consumption growth. The former departs from the aggregate consumption growth ($\bar{G_t}$) based upon the capital share and excess volatility, as follows:\footnote{The stock market participation rate is about 50\% \citep{lettau2019capital}. We assumes that the low income stockholders have the aggregate consumption growth volatility. The consumption growth volatility of the low income stockholders is, therefore, higher than that of labour workers and lower than that of high income stockholders. Excess volatility of high income stockholder consumption growth is absorbed by labour workers and does not affect $\bar G_t$.} 
\begin{equation}
    G^H_{t}=\bar{G_t}f^H_{KS,t}(1+\xi_t)
    \label{eq:high_g}
\end{equation}
\begin{equation}
    G^L_{t}=\bar{G_t}f^L_{KS,t}
    \label{eq:low_g}
\end{equation}
where the stochastic term $\xi_t\sim N_{i.i.d}(0,\Sigma)$ captures the excess volatility of the consumption growth rate of the high income group compared to the low income group.\footnote{Our model does not make an autocorrelation assumption for $\xi_t$ to avoid possible explosive growth. $\Sigma$ denotes a constant variance for $\xi_t$.} The consumption growth volatility shocks of the high income stockholders are absorbed by the labour workers, and the aggregate consumption growth $\bar{G_t}$ remains independent from such shocks according to equation (\ref{EQ:G_OVERALL}). The $\xi_t$ term therefore defines the variance of the consumption distribution of the economy. Based on the data correlations identified by \cite{lettau2019capital}, we formulate $f^H_{KS,t}$ as a monotonic increasing function, and $f^L_{KS,t}$ as a monotonic decreasing function, of the capital share growth rate. The volatility of $G^H_t$ is bounded due to limited resources and productivity growth. If the volatility of $G^L_t$ equals zero, the volatility of aggregate consumption growth $\bar{G_t}$ must also be zero when $f^L_{KS,t}$ is non-zero.

Since labour workers do not participate in the stock market, our model only focuses on the partial equilibrium of stockholders. The average stockholder consumption growth $G^S_t$ can be approximated by the weighted average of high and low income stockholder groups:
\begin{equation}
    G^S_{t}\approx w^H G^H_{t}+ w^L G^L_{t}
    \label{eq:total_g}
\end{equation}
Substituting disaggregate consumption in equations (\ref{eq:high_g}) and (\ref{eq:low_g}) into stockholder consumption growth in equation (\ref{eq:total_g}):
\begin{equation}
    G^S_{t}=\bar{G_t}[w^H f^H_{KS,t}(1+\xi_t)+ w^Lf^L_{KS,t}]
    \label{eq:total_g2}
\end{equation}
Given evidence from \cite{saez2016wealth}, \cite{lettau2019capital} assume that the stockholder consumption equals to the product of the aggregate consumption and capital share. Inspired by \cite{lettau2019capital}, we assume that high (low) income stockholders' consumption growth is positively (negatively) related to the capital share growth, such that  $f^H_{KS,t}=1+F_{KS,t}$ and $f^L_{KS,t}=1-F_{KS,t}$.\footnote{We define capital share growth as capital share factor $F_{KS,t}$ to be consistent with the notation used by \cite{lettau2019capital}.}\footnote{The empirical evidence presented by \cite{lettau2019capital} points out that bottom 90\% wealth distribution is strongly negatively correlated with capital share growth. Using \cite{saez2016wealth} data, \cite{lettau2019capital} Table 2 identifies that capital share has a negative and statistically significant resource impact upon low income U.S. stock owners (OLS coefficient = -1.27, t-statistic = - 6.82). In comparison capital share has a positive resources impact upon high income stock owners which is approximately equal and opposite (OLS coefficient = 1.20, t-statistic = 7.34).} Therefore, the consumption of each stockholder group contains a persistent component as in \cite{bansal2004risks}.\footnote{We assume two stockholder groups and each group satisfies assumptions of \cite{bansal2004risks} independently.}

Thus, stockholder consumption growth in equation (\ref{eq:total_g}) can be written as:\footnote{In equation (\ref{eq:total_g4}), $w^H \in [0, 1]$ by definition. About 95\% of $F_{KS}$ falls in the range between -4\% and 4\% (within two standard deviations) as shown by the empirical results in Table \ref{tab:kssq_des}. The bounded volatility of $C^H_t$ also implies a bounded excess volatility $\xi_t$.  Therefore, the $\bar{G_t}[w^H(1+ F_{KS,t})]\xi_t$ term is bounded and the stockholder consumption does not witness an explosive growth in this model.}
\begin{equation}
      G^S_{t}=\bar{G_t}[w^H(1+ F_{KS,t})+ w^L(1-F_{KS,t})]+\bar{G_t}[w^H(1+ F_{KS,t})]\xi_t
       \label{eq:total_g4}
\end{equation}
Any percentage at the top can be used to illustrate how the concentration of wealth affects the intensive margin of the stock market  \citep{lettau2019capital}.\footnote{The wealth-weighted participation rate is lower than the aggregate participation rate, regardless which quantile of wealth distribution is selected as a benchmark.} We assume that stockholder wealth and population wealth in the economy are 
drawn from the same distribution, to solve the population sizes of each stockholder income group. Our model therefore approximates the high income stockholder population weight $w^H$ by the stock market participation rate ($w^H\approx r^p$), and low income stockholder population weight $w^L$ by labour worker population weight in the economy ($w^H\approx1-r^p$). Given the average stock market participation rate is close to 50\% over time \citep{lettau2001resurrecting}, we can simply assume $w^H = w^L$ for simplicity of the calculation.

Rewriting aggregate consumption growth in equation (\ref{eq:total_g4}), given a relatively small $F_{KS,t}$ and the fact that $w^H-w^L= 0$, the positive impact of capital share growth on the high income group and the negative impact of capital share growth cancel one another out. At time $t$, taking conditional expectations of (\ref{eq:total_g4}) and, therefore, setting the stochastic term to zero, we have the expected stockholder consumption growth as:
\begin{align}
      E_t[G^S_{t}]&=E_t[\bar{G_t}(w^H(1+ F_{KS,t})+ w^L(1-F_{KS,t}))]\nonumber\\
      &=\bar{G_t}[1+ E_t(F_{KS,t})(w^H-w^L)]\nonumber\\
                &= \bar{G_t}
                \label{eq:expected_growth}
\end{align}
Notice that in equation (\ref{eq:expected_growth}), when the dynamics of low and high income stockholder consumption growth $f^L_{KS,t}$ and $f^H_{KS,t}$ have different functional forms, their mutual effect on the level of expected stockholder consumption growth can be tested can be tested by having the capital share factor in the excess return equation. In contrast, equation (\ref{eq:total_g4}) contains the $\bar{G_t}[w^H(1+ F_{KS,t})]\xi_t$ term, which indicates the stockholder consumption growth volatility operates through capital share growth. The magnitude of aggregate stockholder consumption growth volatility also associates with the population size of the high income group $w^H$, and the excess volatility $\xi_t$. Equations (\ref{eq:total_g4}) and (\ref{eq:expected_growth}) are consistent with the empirical findings of \cite{lettau2019capital} in that the consumption growth of the top wealth distribution is more volatile than that of the rest of the population, but the expected consumption growth rate of the richest individuals is at around the same level as that of the whole economy.

To model stockholder consumption growth, we further assume the aggregate consumption growth rate $g_t$ contains the persistent expected growth rate component $x_t$ proved by \cite{bansal2004risks} to define the centre of the consumption growth distribution for the economy. We define the aggregate consumption growth rate $g_{t+1}=log\bar{G}_{t+1}=\mu+x_{t}+\sigma\eta_{t+1}$, and $x_{t}$ is the predictable term following \cite{bansal2004risks}.\footnote{$\bar{G}_{t+1}=1+g_{t+1}$} According to the dynamics of stockholder consumption growth described by equation (\ref{eq:total_g4}) and the aggregate consumption growth $\bar{G}$, the stockholder consumption growth rate $g_{t+1}$ can be written as the following function:\footnote{The $w^H(1+F_{KS,t+1})\xi_{t+1}$ term is relatively small given the range of $w^H$ and $F_{KS,t+1}$, and the definition of $\xi_{t+1}$. Therefore, we use a Taylor approximation here.}
\begin{equation}
     g_{t+1}=\mu+x_{t}+w^H(1+F_{KS,t+1})\xi_{t+1}+\sigma\eta_{t+1}
    \label{eq:log_g_apx}
\end{equation}
Therefore, the time-varying volatility of stockholder consumption growth rate $\sigma_{t+1}$ is defined as $w^H(1+F_{KS,t+1})\xi_{t+1}+\sigma\eta_{t+1}$.

The intuition behind the consumption growth rate in equation (\ref{eq:log_g_apx}) is as follows.  According to \cite{saez2016wealth}, wealth is highly concentrated at the top of the wealth distribution. For example, the richest 5\% of the population owns more than half of the aggregate wealth in the economy. Capital share change does not affect the conditional expectation of stock owner consumption growth, see equation (\ref{eq:expected_growth}). However, when the capital share growth is positive, the wealth increase of individuals in the high income stockholders is higher than in the low income stockholders, as addressed by \cite{gabaix2016dynamics}.\footnote{This statement does not conflict with assumed high and low income stockholder consumption growth in equations (\ref{eq:high_g}) and (\ref{eq:low_g}): the aggregate consumption $\bar{G_t}$ is also increased when capital share increases, as they are positively correlated.} The consumption of the top of the wealth distribution will have a larger impact on aggregate consumption growth when capital share increases. Therefore, stockholder consumption growth volatility is positively correlated with capital share growth.

Also relevant to our model is the consumption volatility risk (CVR) factor derived by \cite{boguth2013consumption}.
In the theoretical motivation of their volatility risk factor, the consumption growth is assumed to switch between high and low volatility states. In our model, instead of assuming a Markov switching process based upon changing beliefs, volatility is explicitly modeled using a capital share factor.  
In \cite{boguth2013consumption}, consumption growth volatility $\sigma_t$ is assumed to be a time varying function of high ($\sigma^H$) and low ($\sigma^L$) volatility states:
 \begin{equation}
 \hat{\sigma_t}=b_t\sigma^H+(1-b_t)\sigma^L=b_t(\sigma^H-\sigma^L)+\sigma^L
     \label{eq:cvr}
 \end{equation}
 In our model, according to the innovation of consumption growth in equation (\ref{eq:log_g_apx}), the volatility of consumption growth is assumed to be:
 \begin{equation}
     \hat{\sigma_t}=w^H(1+F_{KS,t+1})\xi_{t}+\sigma\eta_{t}
     \label{eq:ourvol}
 \end{equation}
The excess volatility of the high state to the low state ($\sigma^H-\sigma^L$) and $\sigma^L$ in equation (\ref{eq:cvr}) are explained in our model as excess volatility of the high to the low income stockholders ($\xi_t$) and $\sigma\eta_{t+1}$ in equation (\ref{eq:ourvol}) respectively. Therefore, in equation (\ref{eq:ourvol}), the counterpart of belief $b_t$ is $w^H(1+F_{KS,t+1})$, see equation (\ref{eq:cvr}).

\subsection{A Model of Equity Returns}

In line with the assumption by \cite{lettau2019capital}, the labour workers do not influence equity prices and, consequently, they are independent from the stock market and their participation is not modeled. Equity returns are  linear functions of the equity premium according to the Capital Asset Pricing Model \citep{cochrane2009asset}. To solve the relationship between equity returns and capital share growth, our paper extends the model of \cite{bansal2004risks} to derive the equity premium explicitly. The system stated in our paper is a hybrid system of the constant volatility case (Case I) and the time-varying volatility case (Case II) of \cite{bansal2004risks}. The volatility of the log stockholder consumption growth contains both a constant element $\sigma$ and a time-varying part $w^H(1+F_{KS,t+1})\xi_{t+1}$. The dividend growth volatility is correlated with consumption growth volatility, as suggested by \cite{bansal2004risks}. Therefore, $\sigma_{d,t+1}$ is assumed to be partially correlated with both $F_{KS,t+1}\xi_{t+1}$ and $\sigma$ in our model.\footnote{The specification of $\sigma_{d,t+1}$ also relaxes the setting by \cite{bansal2004risks} Case II, in which $g_{t+1}$ and $g_{d,t+1}$ are cointergated, to be consistent with empirical literature \citep{campbell1999force}. } 

In our model, the stock market is driven by a persistent growth component ($x_{t+1}$), capital share and stochastic high income volatility shocks ($\xi_{t+1}$) based upon equation (\ref{eq:log_g_apx}) as follows:  
\begin{align}
    &x_{t+1}=\rho  x_{t}+\phi_e\sigma e_{t+1}\nonumber\\
    &g_{t+1}=\mu+x_{t}+w^H(1+F_{KS,t+1})\xi_{t+1}+\sigma\eta_{t+1}\nonumber\\
    &g_{d,t+1}=\mu_d+\phi x_{t}+\phi_d\sigma_{d,t+1} u_{t+1}  \label{eq:system_g}\\
    & ~~~~~~~~~~e_{t+1},u_{t+1},\eta_{t+1}\sim N_{i.i.d.}(0,1)~~~~~~\xi_{t}\sim N_{i.i.d.}(0,\Sigma)\nonumber
\end{align}
where the $g_{d,t+1}$ is the log dividend growth rate, and $\rho$ is the persistence of the expected growth rate process. Parameters $\mu$ and $\mu_d$ are the constant component of $g_{t+1}$ and $g_{d,t+1}$, respectively. $\phi_e>1$ and $\phi_d>1$ allow for parameter calibration. The parameter $\phi$ can be interpreted as the leverage ratio
on expected consumption growth, see \cite{bansal2004risks} and \cite{abel1999risk}. The stochastic error terms $e_{t+1}$, $u_{t+1}$, and $\eta_{t+1}$ are independent from each other \citep{bansal2004risks}. $\sigma$ is a constant which captures the volatility of $x_{t+1}$ and $g_{t+1}$.\footnote{\cite{bansal2004risks} Case II adds time varying volatility and fluctuating economic uncertainty into their model through a general error term. Our model does not assume a stochastic innovation of $\sigma$ in order to isolate the volatility effect generalized by the introduction of the capital share factor.}  The innovation of $g_{d,t+1}$ which is found to be more volatile than $g_{t+1}$ \citep{campbell1999asset} is tackled by $\phi_d$. Our model therefore formalises uncertainty in terms of the impact of high income consumption variability, rather than a generic uncertainty as set out by \cite{bansal2004risks}.

Based upon the recursive preference utility function, the asset pricing restrictions for gross return $R_{i,t+1}$ satisfy
 \begin{equation}
     E_t[\delta^\theta G^{-\frac{\theta}{\psi}}_{t+1}R^{-(1-\theta)}_{a,t+1}R_{i,t+1}]=1
     \label{eq:recursive}
 \end{equation}
where $\theta=(1-\gamma)/(1-\frac{1}{\psi})$. In equation (\ref{eq:recursive}), $G_{t+1}$ denotes the aggregate consumption growth rate, and $R_{a,t+1}$ is the gross return on an asset that generates dividends that cover the aggregate stockholder consumption. $0<\delta<1$ is the time discount factor, $\gamma\geq0$ is the risk-aversion parameter, and $\psi\geq0$ is the intertemporal elasticity of substitution (IES). 

Given the asset pricing constraint in equation (\ref{eq:recursive}), the intertemporal marginal rate of substitution (IMRS):
\begin{equation}
    m_{t+1}=\theta log\delta-\frac{\theta}{\psi}g_{t+1}+(\theta-1)r_{a,t+1}
    \label{eq:imrs}
\end{equation}
where $g_{t+1}$ and $r_{a,t+1}$ are the natural logarithm of $G_{t+1}$ and $R_{a,t+1}$, respectively.

We also adopt the standard approximation proposed by \cite{campbell1988stock} to derive the functional form of the equity premium. The innovation of log gross consumption $r_{a,t+1}$ and log market return $r_{m,t+1}$ are assumed to follow:
\begin{equation}
    r_{a,t+1}=\kappa_0+\kappa_1 z_{t+1}-z_t+g_{t+1}
    \label{eq:ra_apx}
\end{equation}
\begin{equation}
    r_{m,t+1}=\kappa_{0,m}+\kappa_{1,m} z_{m,t+1}-z_{m,t}+g_{d,t+1}
    \label{eq:rm_apx}
\end{equation}
where $z_t$ is the log price-consumption ratio ($log(\frac{P_t}{C_t})$) and $z_{m,t}$ is the log price-dividend ratio ($log(\frac{P_t}{D_t})$).\footnote{$D_t$ denotes the dividend.} Therefore, $z_t$ and $z_{m,t}$ are assumed to satisfy $z_t=A_0+A_1x_t+A_{2,t}\xi_t$ and $z_{m,t}=A_{m,0}+A_{m,1}x_t+A_{2,m,t}\xi_t$.\footnote{$A_0$ and $A_{m,0}$ are constants; $A_1$ and $A_{m,1}$ are parameters of the persistent consumption growth component $x_t$; $A_{2,t}$ and $A_{2,m,t}$ are parameters of the excess volatility $\xi_t$} The relevant state variables in solving for the equilibrium are $x_t$ and $\xi_t$. We modify the functional form of the log price-consumption and log price-dividend ratios assumed by \cite{bansal2004risks} to include the time-varying part of stockholder consumption growth volatility.\footnote{ See \cite{bansal2004risks} Case II.} 

In addition, we need the dynamics of capital share growth to solve the equity premium. According to \cite{lettau2019capital}, the capital share growth follows an AR(1) process:\footnote{The constant is not significant according to our AR(1) estimation.} 
\begin{equation}
    F_{KS,t+1}=\rho^{KS}F_{KS,t}+e^{KS}_{t+1}
    \label{eq:ks_growth}
\end{equation}
where $e^{KS}_{t+1}$ captures unexpected shocks in capital share growth. 

Since the log consumption growth $g_{t}$, log dividends growth $g_{d,t}$ , and the capital share growth are exogenous processes in our system, the functional form of the innovation of consumption return, the pricing kernel, and equity returns in this economy can be derived explicitly using equations (\ref{eq:imrs})- (\ref{eq:ks_growth}).\footnote{Detailed proofs are provided in the Appendix.}

We first solve the parameters of the persistent consumption growth $x_t$ and excess volatility $\xi_t$ on price-consumption and price-dividend ratios, which track expected risk prices \citep{campbell2000explaining}. In our model, the resulting $A_{1}$ and $A_{1,m}$ are identical to \cite{bansal2004risks}. The sensitivity of the price-consumption (and price-dividend) ratio to the excess volatility $\xi_{t}$ is constant over time. $A_{2,t}$ (and $A_{2,m,t}$) are constants when we hold $w^H$, $\rho_{KS}$ and $F_{KS,t}$ constant:\footnote{$A_{2,t}$ and $A_{2,m,t}$ are derived in Appendix.} 
\begin{align}
    &A_{2,t}=\frac{1-\frac{1}{\psi}}{1-\kappa_1}w^H \rho_{KS}F_{KS,t}\label{eq:A_2,t}\\
    &A_{2,m,t}=\frac{\theta-1-\frac{\theta}{\psi}}{1-\kappa_{1,m}}w^H -\frac{w^H\rho_{KS}}{\psi(1-\kappa_{1,m})}F_{KS,t}\label{eq:A_2,m,t}
\end{align}
According to the parameters of excess volatility in equations (\ref{eq:A_2,t}) and (\ref{eq:A_2,m,t}), given the stochastic nature of $\xi_{t}$, the capital share growth does not have an impact on the magnitude but affects the uncertainty of the price-consumption and price-dividend ratios. Therefore, due to constant excess volatility between two adjacent periods, capital share growth does not shift the expected rate of return under short-run (conditional) expectations. However, in the long-run, volatility shocks fail to feature in expectations and the increased uncertainty of returns generate redistribution risks between high and low income stockholders.

We now set out equity returns conditionally and unconditionally. The difference between these two settings is due to the difference between conditional and unconditional expectations of $\xi_{t+1}$. Conditioning on information at $t$, $E_t(\xi_{t+1})=\xi_t$ due to smoothed consumption, while the unconditional expectation of $\xi_{t+1}$ is $0$. 
\subsection{Conditional and Unconditional Expectations and Equity Premiums}
We now derive equity premiums under conditional and unconditional expectations, respectively. \footnote{Full details are in the Appendix.}

Conditional on information at time $t$, all shocks alter agents' expectations. The conditional innovation of the pricing kernel $m_{t+1}$ is:
\begin{align}
      m_{t+1}-E_t(m_{t+1})=\lambda_{\eta} \sigma \eta_{t+1}+\lambda_{e}  \sigma e_{t+1}+\lambda_{\xi,t+1} \xi_{t+1}
    \label{eq:price_kernel}  
\end{align}
The conditional innovation of market return $r_{m,t+1}$ is: 
\begin{align}
r_{m,t+1}-E_t(r_{m,t+1}) =\phi_d\sigma_{d,t+1} u_{t+1}+\lambda_{m,e} \sigma e_{t+1}+\lambda_{m,\xi,t+1} \xi_{t+1}\label{eq:c_rm_inno}
\end{align}
In equations (\ref{eq:price_kernel}) and (\ref{eq:c_rm_inno}), $\lambda_{m,e}$, $\lambda_{\eta}$ and $\lambda_{e}$ are constants, while $\lambda_{m,\xi,t+1}$ and $\lambda_{\xi,t+1}$ are functions of $e^{KS}_{t+1}$.\footnote{See the Appendix.} Therefore, the conditional pricing kernel innovation in equation (\ref{eq:m_inno2}) is only correlated to unexpected capital share growth $e^{KS}_t$, but the conditional market return innovation is correlated with capital share growth through $\sigma_{d,t+1}$.

Following \cite{bansal2004risks}, the continuous equity premium in the presence of time-varying economic uncertainty is
\begin{align}
    E_t(r_{m,t+1}-r_{f,t})=&-(\lambda_{m,e}\lambda_{e}-0.5\lambda^2_{m,e})\sigma^2+0.5\phi^2_d\sigma^2_{d,t+1}+E_t(\lambda_{m,\xi,t+1}\lambda_{\xi,t+1}-0.5\lambda^2_{m,\xi,t+1})\nonumber\\
    =&-(\lambda_{m,e}\lambda_{e}-0.5\lambda^2_{m,e})\sigma^2+0.5\phi^2_d\sigma^2_{d,t+1}
    \label{eq:tvequitypremium}
    \end{align}
At time $t$, the conditional expectation $E_t(\xi_{t+1})=\xi_t$, so the effect of predictable capital share growth is omitted in equation (\ref{eq:tvequitypremium}). As shown by equation (\ref{eq:tvequitypremium}), the conditional equity premium is constant and has one source of systematic risk that relates to fluctuations in expected consumption growth $\sigma^2$. However, the capital share factor enters the innovation of market return in equation (\ref{eq:c_rm_inno}). Hence, the excess volatility of the high income group, through capital share growth, is linked to the variability of equity returns. We estimate the conditional equity premium in equation (\ref{eq:tvequitypremium}) using the short-window regression as suggested by \cite{lewellen2006conditional} and the Bayesian approach proposed by \cite{bianchi2017macroeconomic}.

Under unconditional expectations, we do not allow unexpected shocks of parameters. The unconditional innovation of the pricing kernel is as follows:
\begin{align}
    m_{t+1}-E(m_{t+1})=\lambda_{\eta} \sigma \eta_{t+1}+\lambda_{e}  \sigma e_{t+1}+\lambda^u_{\xi,t+1} \xi_{t+1}\label{eq:m_inno2}
\end{align}
The unconditional innovation of market return is:
\begin{align}
     r_{m,t+1}-E(r_{m,t+1})= \phi_d\sigma u_{t+1}+\lambda_{m,e} \sigma e_{t+1}+\lambda^u_{m,\xi,t+1} \xi_{t+1}\label{eq:uc_rm_inno2}
\end{align}

Detailed functional forms of the parameters in equations (\ref{eq:m_inno2}) and (\ref{eq:uc_rm_inno2}) are in the Appendix.
Using equations (\ref{eq:m_inno2}) and (\ref{eq:uc_rm_inno2}), the unconditional equity premium is calculated as: 
\begin{align}
    E(r_{m,t+1}-r_{f,t})=-(\lambda_{m,e}\lambda_{e}-0.5\lambda^2_{m,e}-0.5\phi^2_d)\sigma^2+E[\lambda^u_{m,\xi,t+1}\lambda^u_{\xi,t+1}-0.5(\lambda^u_{m,\xi,t+1})^2]
    \label{eq:tvequitypremium2}
    \end{align}
where $E[\lambda^u_{r,\xi,t+1}\lambda^u_{\xi,t+1}-0.5(\lambda^u_{r,\xi,t+1})^2]$ is positively correlated with $E(F^2_{KS,t+1})$.\footnote{See the Appendix.} Under unconditional expectations, the equity premium is a function of fluctuations in expected consumption growth $\sigma^2$ and capital share variability $E(F^2_{KS,t+1})$. 

The intuition behind $E(F^2_{KS,t+1})$ as an unconditional risk factor is as follows. The variance of consumption volatility, captured by $e^{KS}_t$ in our model, is very small and gets magnified under unconditional expectations because of the long-lasting nature of the volatility shock \citep{bansal2004risks}. Intuitively, the ratio of the conditional risk premium to the conditional volatility of the market portfolio fluctuates with consumption volatility \citep{bansal2004risks}. The maximal Sharpe ratio approximated by volatility of the pricing kernel innovation also varies with consumption volatility. In our model, consumption volatility operates through capital share growth. Therefore, risk prices will rise as economic uncertainty represented by capital share variability rises. Conditional on both of the two stockholder groups surviving in the long-run, the magnitude of IES, $\psi$, is justified by the survival analysis of \cite{borovivcka2020survival} which studies a two-agent model from the perspective of beliefs, where under different belief styles, IES is found to be greater than 1 to ensure the long-run coexistence of two heterogeneous agents. When $\psi>1$ holds, the negative coefficient of capital share growth in parameters $A_{2,t}$ in equation (\ref{eq:A_2,t}) and $A_{2,m,t}$ in equation (\ref{eq:A_2,m,t}) ensures that  capital share growth is negatively correlated with the uncertainty in the price-consumption and the price-dividend ratio. In response to lower expected expected rates of return uncertainty, asset demand rises to generate positive risk price of the capital share variability in our model. The utility study of \cite{colacito2018volatility} also highlights that increased macroeconomic volatility increases the stochastic discount factor under the recursive utility framework, thus raises expected returns and generates a positive volatility risk price. We estimate the unconditional equity premium in equation (\ref{eq:tvequitypremium2}) using a  Fama-MacBeth approach suggested by \cite{lettau2019capital} and capital share variability as a risk factor. Given our theoretical framework, the capital share variability risk price is expected to be positive.

The model specification of Boguth and Kuehn's  (\citeyear{boguth2013consumption}) consumption volatility risk (CVR) factor is a potential explanation of the nonlinear relationship between the equity premium and capital share growth in equation (\ref{eq:tvequitypremium2}) due to the following reasons. Changes in beliefs about consumption growth volatility are found important in explaining unconditional equity returns by \cite{boguth2013consumption}, which indicates that the assumption of two volatility states is reasonable. Although the functional forms of equations (\ref{eq:cvr}) and  (\ref{eq:ourvol}) are similar, the change of volatility in equation (\ref{eq:ourvol}) is a smoothed process. Thus, our model can be alternatively explained by assuming infinite states of consumption growth volatility. At each time $t$, consumption growth volatility has only two latent states, but $\xi_t$ is an unknown stochastic variable and, hence, this is a setup that is consistent with the quadratic relationship between equity returns and capital share growth in our model.

To conclude, our theoretical model indicates that: under conditional expectations, the capital share factor captures the impact of consumption volatility from the high income group onto equity returns; while under unconditional expectations, capital share variability serves as a risk factor that
captures long-run market volatility.

%% file: chapters/Methodology/Methodology.tex
\section{Econometric Methodology}
We employ both unconditional and conditional estimation approaches to examine the empirical importance of capital share risks and to test the predictions of our model. The unconditional estimation is a bootstrapped \cite{fama1973risk} procedure, which corrects both cross-sectional correlations and the firm effect of equity returns. Conditional estimations include Lewellen and
Nagel's (\citeyear{lewellen2006conditional}) rolling-window regressions, the Bayesian time-varying beta with stochastic volatility (B-TVB-SV) estimation from \cite{bianchi2017macroeconomic}, and a rolling-window multiplicative GARCH. The rolling-window and the B-TVB-SV approaches assume that a risk factor enters the mean equation of the stochastic discount factor as shown in equation (\ref{eq:ks_apx}) to test the explanatory power of capital share growth on the level of equity returns. The rolling-window and the B-TVB-SV estimates are expected to generate statistically insignificant capital share factor loadings according to the conditional stochastic discount factor in equation (\ref{eq:price_kernel}) and the conditional equity premium in equation (\ref{eq:tvequitypremium}). The rolling-window multiplicative GARCH assumes that the capital share factor influences market volatility only, and is used to test the conditional innovation of market return in equation (\ref{eq:c_rm_inno}). 
\input{chapters/Methodology/f-mb-boots.tex}
\subsection{Conditional Cross-sectional Regressions}

\input{chapters/Methodology/RollingWindow.tex}
\input{chapters/Methodology/B-TVB-SV.tex}

%% file: chapters/Methodology/f-mb-boots.tex
\subsection{Unconditional Cross Sectional Regressions}
The risk price measures the risk-reward relationship between factors and returns. Fama and MacBeth's (FMB) two-step procedure is widely used in estimating risk prices of factors and in testing asset pricing models when risk factors enter the mean equation of equity returns. In practice, the static or static based F-MB approach estimates both the mean effect and the variance effect of risk factors together. When a risk factor enters the mean equation of the true data dynamics, which is consistent with the assumption of the static F-MB approach, the risk price estimate of this factor will be significant. However, when a risk factor enters the variance equation of the true dynamics, the ordinary least squares (OLS) regression in the first step of the static F-MB approach will be biased due to heteroskedasticity problems. The risk price estimate of this factor will also be significant because of the change in the width of the factor loading distribution.

Our paper employs the F-MB bootstrap to test the pricing power of the capital share factor and estimate the unconditional equity premium of the capital share variability factor in equation (\ref{eq:tvequitypremium2}). The F-MB bootstrap is based upon the static F-MB procedure, and can be used to correct both cross-sectional correlations and firm effects \citep{lettau2019capital} while it constraints the factor loadings to be constant over time as the static F-MB estimators. 

When testing the importance of the capital share factor for U.S. asset returns, \cite{lettau2019capital} adopt the non-overlapping block residual bootstrap for both steps of the F-MB procedure. Although it is argued that utilizing the overlapping bootstrap is a more robust method, \cite{andrews2004block} compares overlapping and non-overlapping block bootstraps, and reaches the conclusion that although the former is often favored in applications, the latter generates similar numerical results. Our paper uses the non-overlapping bootstrap and the capital share dynamics assumed by \cite{lettau2019capital} for the capital share factor.\footnote{In the first step time series regression, capital share growth is assumed to follow an AR(1) process to factor in the serial correlation.}
 
The optimal length of the bootstrap block should increase as the sample size increases to maintain the consistency of moments and distribution functions \citep{horowitz1997bootstrap}. In the first step, our sample spans 526 months.\footnote{The bootstrap only estimate January 1974 to August 2018  which is consistent with sample span for B-TVB-SV approach.}.Therefore, the optimal block-length is $536^{(\frac{1}{5})}\approx 4$ following \cite{hall1995blocking}. The second step involves 25 portfolio returns, and the optimal block-length is identical to \cite{lettau2019capital}. 

The F-MB bootstrap tackles both the cross-sectional correlation and serial correlation in estimation, but it can only serve as a rough check of pricing power of the capital share factor and cannot infer the functional form of the true dynamics of equity returns. When assuming that capital share growth is a risk factor in the mean equation, the time variation of factor loadings are not captured by simply estimating one time series regression in the first step of F-MB bootstrap approach, hence the risk prices are estimated unconditionally. Additionally, the non-linearity in the dynamics of equity returns is omitted, leading to biased F-MB bootstrap estimates of the capital share risk price. 

According to the theoretical justification in equation (\ref{eq:c_rm_inno}), the capital share factor explains the variance of equity returns under conditional expectations. Also, the true unconditional risk factor should be the capital share variability according to equation (\ref{eq:tvequitypremium2}). In our paper, the monthly F-MB bootstrap estimates are benchmarks of pricing power. The monthly risk price of the capital share factor is expected to be significant due to its multicollinearity with capital share variability $E(F^2_{KS,t})$ in a single capital share factor model.\footnote{$F_{KS,t}$ and $E(F_{KS,t})$ all contains the mean of $F_{KS,t}$ plus terms containing deviation from the mean. Therefore, $F_{KS}$ and $E(F_{KS,t})$ are correlated. } In a two factor model including both the capital share factor and capital share variability, the latter can be expected to dominate according to the unconditional equity premium in equation (\ref{eq:tvequitypremium2}). 

%% file: chapters/Methodology/RollingWindow.tex
\subsubsection{Rolling-Window Fama-MacBeth}
As shown by equation (\ref{eq:tvequitypremium}), the mean equation of the equity premium is independent from the capital share factor. We adopt rolling-window regressions to estimate factor loadings in a conditional manner as suggested by \cite{lewellen2006conditional}. Our paper estimates the F-MB first step regression following \cite{lewellen2006conditional}. The window length selected in the first step F-MB is 12 months. The second step is identical to the original cross-sectional regression of the F-MB approach. The results of the rolling-window regression serves as a benchmark for the true DGP of factor loadings under the assumption of a modest level of temporal variation. 

A short window for estimation is adopted for the following reasons. Within each window, the regression using short horizon data can be viewed as an estimation that is robust to firm effects, especially since the autocorrelation of stock returns is weaker over a relatively short regression window \citep{fama1988permanent}. Another function of the rolling-window regression is to serve as a volatility estimator. Volatility is constant within each window, but varies across windows.  

The limitations of the rolling-window approach are widely known. The rolling-window F-MB is an appropriate approximation for time-varying factor loadings, only conditional on the assumption that there are no structural breaks present within each window. The time variations are still not fully captured due to the ad-hoc window length selection: robustness of the rolling-window approach is diminished when extreme outliers are present in the sample. Therefore, the assumption of rolling-window F-MB is still too strong and vulnerable. Further, the rolling-window F-MB is subject to a common problem of 2-step estimations, which is that the second step estimation is dependant on the first step results. This approach cannot pass the variability of factor loadings into the second step estimation and, therefore, is insufficient to ensure unbiased estimation of risk prices. The rolling-window approach also views the factor loading as a constant at each time point, causing information carried by the change of factor loading volatilities to be retained within the first step estimation. The time variation of risk prices are thus inflated compared to the true underlying DGP by the rolling window F-MB approach when stochastic volatility is present in factor loadings.

As shown by the innovation of market premium in equation (\ref{eq:tvequitypremium}), the loading of the capital share factor is expected to be centered at zero, and a strong volatility clustering is expected to be present under rolling-window estimation. Due to heteroskedasticity and the model misspecification problem highlighted by \cite{jagannathan1996conditional}, risk price estimates should be insignificant but vary dramatically over time.\footnote{An insignificant risk factor in the true equity dynamic might be significant under F-MB estimations.} 

%% file: chapters/Methodology/B-TVB-SV.tex
\subsubsection{The Bayesian Time-Varying Beta With Stochastic Volatility Model}
To tackle problems in the F-MB procedures, \cite{bianchi2017macroeconomic} proposes a Bayesian estimation approach, namely the Bayesian time-varying beta with stochastic volatility (B-TVB-SV) model, to consider the SDF and non-arbitrage restriction jointly. Compared to the rolling-window F-MB, this method captures the time variation and variability of factor loadings while maintaining robustness to firm effects. 

The B-TVB-SV model for asset return $r_{i,t}$ as a function of risk factor $F_{j,t}$ is:
\begin{equation}
    r_{i,t}=\beta_{i0,t}+\sum^ K_{j=1}\beta_{ij,t}F_{j,t}+\sigma_{i,t}\epsilon_{i,t}~~~~~~~~~~~\epsilon_{i,t}\sim N(0,1)
    \label{eq:btvbsv1}
\end{equation}
Factor risk prices $\lambda_{j,t}$ are estimated by: 
\begin{equation}
    r_{i,t}=\lambda_{0,t}+\sum^ K_{j=1}\lambda_{j,t}\beta_{ij,t}+e_{i,t}~~~~~~~~~~~e_{i,t}\sim N(0,\tau^2)
    \label{eq:btvbsv2}
\end{equation}
The B-TVB-SV framework assumes the time-varying betas $\beta_{ij,t}$ and residuals in equation (\ref{eq:btvbsv1}) take the following forms:
\begin{equation}
    \beta_{ij,t}=\beta_{ij,t-1}+\kappa_{ij,t}\eta_{ij,t}~~~~~~~~~~~~~j=0,...,K
    \label{eq:tvb}
\end{equation}
\begin{equation}
    ln(\sigma^2_{i,t})=ln(\sigma^2_{i,t-1})+\kappa_{iv,t}\upsilon_{i,t}~~~~~~~~~~~~~i=0,...,N
    \label{eq:sv}
\end{equation}
where $\kappa_{ij,t}$ is the structural break of factor loading $\beta_{ij,t}$, and  $\kappa_{iv,t}$ is the structural break of idiosyncratic variance $ln(\sigma^2_{i,t})$. The stochastic terms $\eta_{ij,t}$ and $\upsilon_{i,t}$ follow normal distributions with zero mean and variances $q^2_{ij}$ and $q^2_{iv}$, respectively. A $\kappa_{ij,t}$ equal to one indicates that structural breaks are present in the factor loadings, and $\kappa_{iv,t}$ equal to one indicates that structural breaks are present in the idiosyncratic variance. The advantage of including structural breaks is that the model captures discrete movements of the factor loadings. In equation (\ref{eq:tvb}), the innovation of factor loading maintains the random walk properties to retain the shrinkage power of the selected prior to the largest extent. Therefore, the B-TVB-SV approach tackles factor selection automatically.\footnote{Weak priors are used for the distributions of $\beta_{ij,t}$ and $ln(\sigma^2_{i,t})$. Evidence indicates when the number of variable is small (K=5), flat prior works quite well with the sparse specification and performs modest with the dense specification \citep{huber2020inducing}. The weak prior adopted by V-TVB-SV approach also has shrinkage effects. } Other detailed break and risk price prior specifications and sampling approaches are discussed in the Appendix.

As shown in equations (\ref{eq:sv}) and (\ref{eq:btvbsv1}), the model specification of the B-TVB-SV estimation allows volatility change to have structural breaks as well as autocorrelations, incorporating variance effects of the risk factors that are assumed to enter the mean equation. Therefore, the B-TVB-SV approach is a robustness check for the true data dynamics of equity returns: with the model specification of the B-TVB-SV, the risk factor will generate a significant factor loading and risk price estimates if the risk factor enters the mean equation of equity returns. Given that the model misspecification problem is corrected by the B-TVB-SV, the distribution of capital share risk price should be centered at zero at each time $t$ as indicated by equation (\ref{eq:tvequitypremium}), and the variance of this distribution should change over time, as in equation (\ref{eq:c_rm_inno}).

\subsection{Rolling-window Multiplicative GARCH}
As shown by the conditional market return innovation in equation (\ref{eq:c_rm_inno}), the capital share should be estimated in the variance equation instead of in the mean equation conditionally. Our paper employs a rolling-window multiplicative GARCH approach to estimate the true volatility effect of the capital share factor on equity returns directly. Within each regression window, the asset pricing model estimated by the multiplicative GARCH is as follows:
\begin{equation}
        r_{i,t}=\beta_{i0,t}+\epsilon_{i,t}~~~~~~~~~~~\epsilon_{i,t}\sim N(0,\sigma^2_{i,t})
    \label{eq:mfapm_kssigma}
\end{equation}
where $Var(\epsilon_{i,t})=\sigma^2_{i,t}$, and $\sigma^2_{i,t}$ is consistent with the form in equation (\ref{eq:garch2}) below. Therefore, the conditional variance is assumed to be correlated with the capital share growth rate.

To test our theoretical predictions in the conditional expectation case, our paper employs the following most general form for the variance equation:
\begin{equation}
    \sigma^2_{i,t}=\gamma_{KS} F^2_{KS,t}
    \label{eq:garch}
 \end{equation} 
The functional form in equation (\ref{eq:garch}) is motivated by the market return innovation. As indicated by equation (\ref{eq:c_rm_inno}), the capital share factor is an $O(n^2)$ addend in the variance equation $\sigma^2_{i,t}$. 
This paper adopts the conditional variance form proposed by \cite{judge1988introduction} to test the variance equation  (\ref{eq:garch}). The capital share factor enters the variance specification as multiplicative heteroskedasticity. Due to the constraint $\sigma^2\geq0$, equation (\ref{eq:garch}) is rewritten as equation (\ref{eq:garch2}) for the sake of estimation.
\begin{equation}
    \sigma^2_{i,t}=exp[\lambda_0+\lambda_1log(F^2_{KS,t})]
   \label{eq:garch2}
\end{equation}
A 60-month window length is selected by this paper for the rolling-window multiplicative GARCH due to the limitation of maximum likelihood convergence. According to the new data dynamics in equations (\ref{eq:mfapm_kssigma}) and (\ref{eq:garch}), the coefficient of $log(F^2_{KS,t})$ in the variance equation (\ref{eq:garch2}) is expected to be significant over time if equations (\ref{eq:c_rm_inno}) and (\ref{eq:tvequitypremium}) hold.

%% file: chapters/4-Data-Priors/4-Data-Prior.tex
\section{Data}
\noindent
\cite{lettau2019capital} use quarterly capital share and quarterly portfolio returns converted from monthly data to test capital share growth. In our paper, instead of modifying monthly returns in a relatively ad-hoc manner, we interpolate the capital share using a reasonable indicator to reduce information loss. 

\input{chapters/4-Data-Priors/RAReturns.tex}

%% file: chapters/4-Data-Priors/RAReturns.tex
\subsection{Capital Share Factor and Variability}
\begin{figure}[ht]
    \centering
     \includegraphics[width=\textwidth]{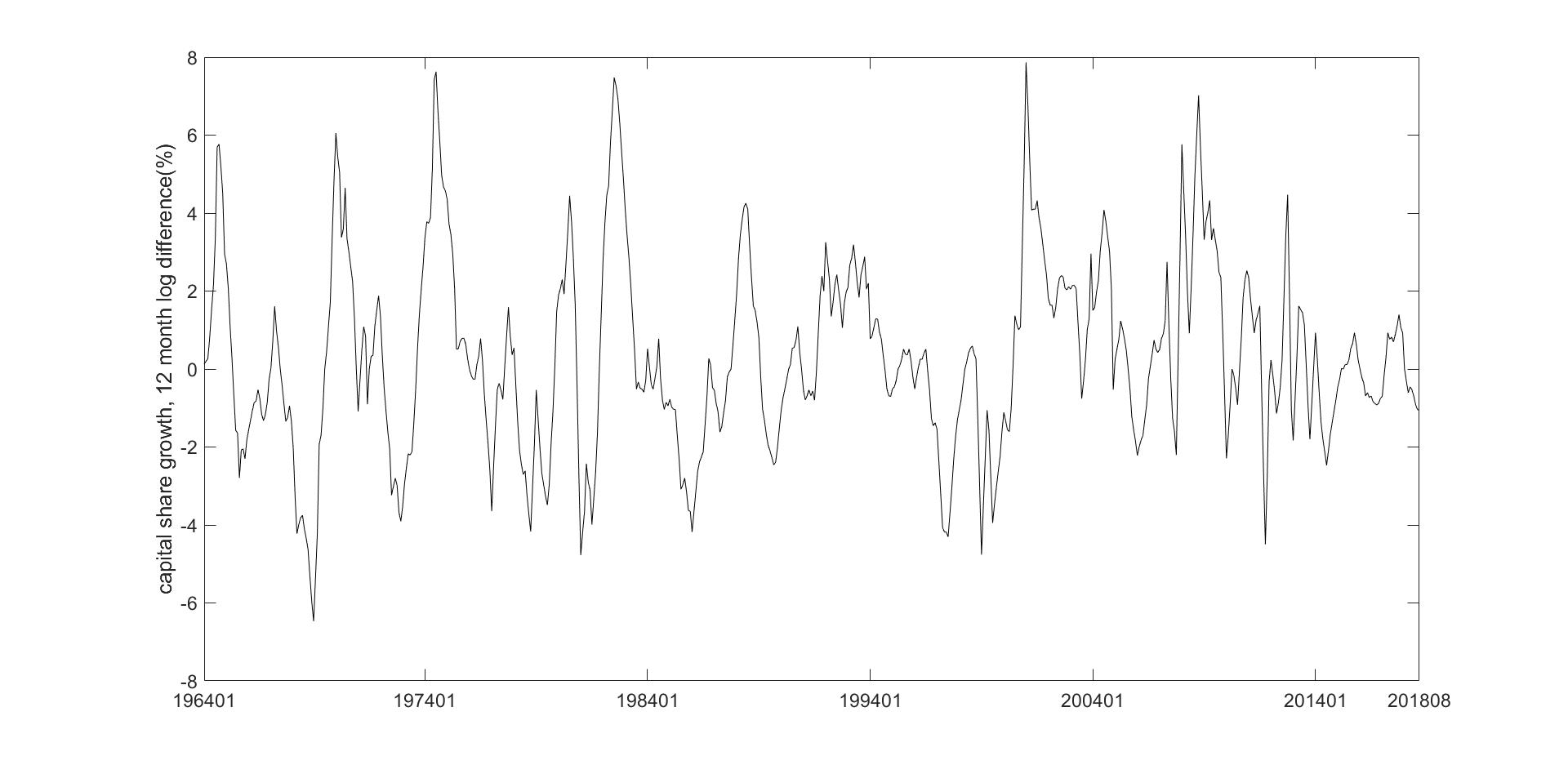}
    \includegraphics[width=\textwidth]{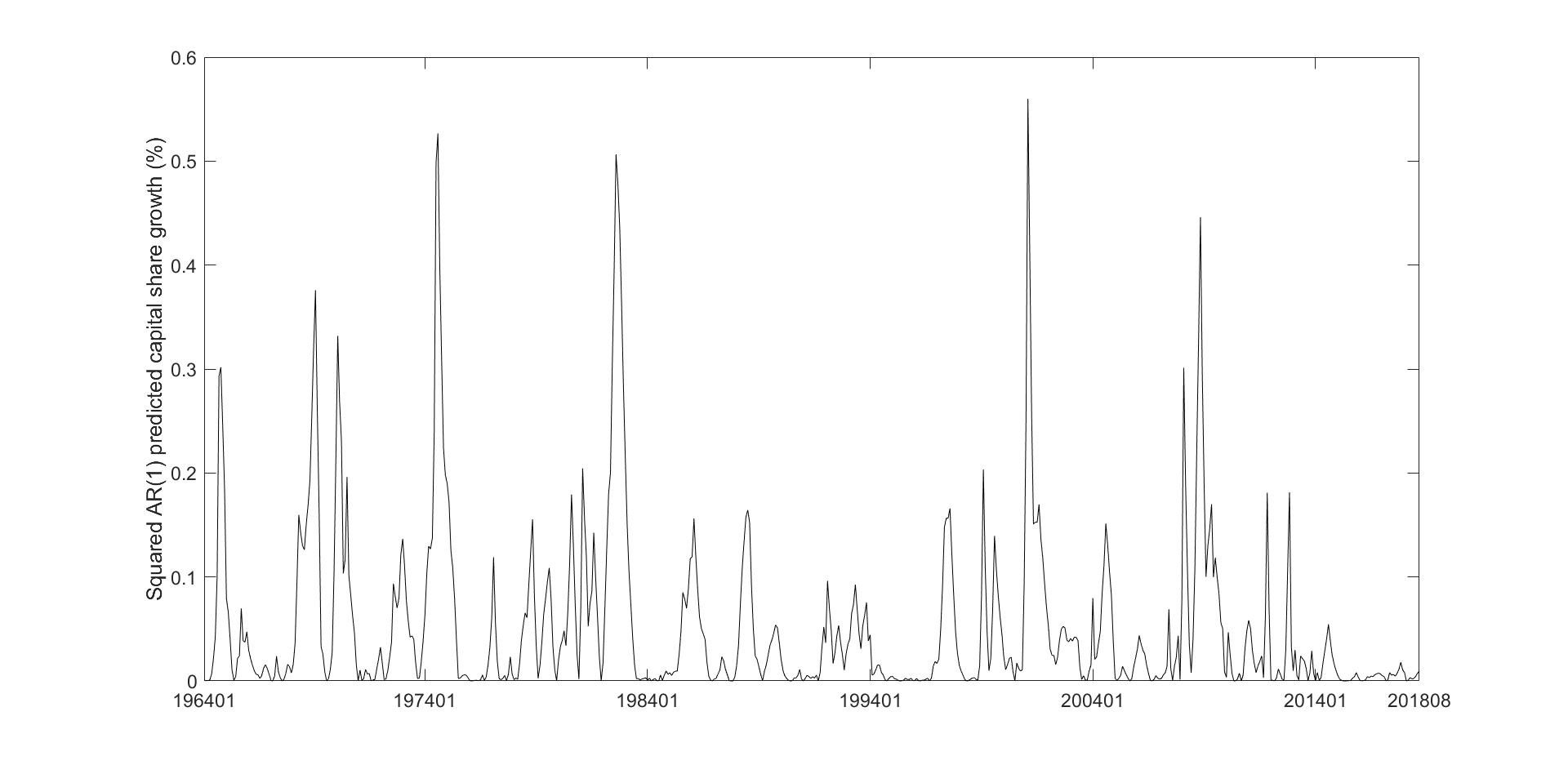}
    \begin{minipage}{.8\linewidth}{
\footnotesize \smallbreak    \caption{\textbf{Capital share growth and variability (\%).} The sample spans January 1964 to August 2018.}
    \label{fig:capitalsharesq}}\end{minipage}
\end{figure}
Measurement error leads to biased estimation of CAPMs \citep{lettau2019capital}.\footnote{During the data collection process, the filtering approach introduces measurement error problem.} Long-term capital share growth is adopted to partial out the measurement error effect. In the test of the capital share factor, \cite{lettau2019capital} compares 1,4,8,12 and 16-quarter capital share growth to tackle measurement error problems. The 4-quarter capital share growth is found to have higher pricing power. 
\smallbreak
Capital share is calculated as $1-Labour~share$. Labour share data is the nonfarm sector labor share, which is identical to that used by \cite{lettau2019capital} and \cite{gomme2004measuring}. Data for constructing capital share from FRED, the monthly capital share is obtained by the Chow-lin interpolation.\footnote{See the Appendix.}
\smallbreak
The original capital share factor (quarterly) constructed by \cite{lettau2019capital} is:
\begin{equation}
    F_{KS}^q=\frac{KS^q_{t+4}}{KS^q_{t}}
    \label{eq:ks_og}
\end{equation}
In equation (\ref{eq:ks_og}) $F_{KS}$ can be decomposed as the capital share growth rate plus a constant 1, indicating that the factor is partially correlated with the intercept. Therefore, the estimated capital share factor loading is higher due to the partial effect taken from the constant. Additionally, the estimated distribution of capital share factor loading tends to be wider due to a higher estimated variance. Finally, from the perspective of the B-TVB-SV, the break probabilities of the capital share factor is not easily identified if the factor is correlated with the constant. To avoid these problems, we use a 12-month capital share growth rate as a risk factor and test its pricing power. The monthly capital share factor tested in this paper is constructed as:
\begin{equation}
    F_{KS}=\frac{KS_{t+12}}{KS_{t}}-1
\end{equation}

According to the unconditional expectation of equity returns in equation (\ref{eq:tvequitypremium2}), capital share variability ($E(F^2_{KS})$) enters the unconditional mean equation of equity returns. The $E(F^2_{KS})$ risk factor is constructed based upon the AR(1) innovation process of $F_{KS}$ as in \cite{lettau2019capital}:
\begin{equation}
    F_{KS,t+1}=\rho^{KS}F_{KS,t}+e^{KS}_{t+1}
\end{equation}
where $e^{KS}_{t+1}$ captures unexpected shocks in capital share growth. The magnitude of the estimate of $\rho^{KS}$ is 0.947, which is statistically significant at 5\% level. We obtain the capital share variability factor using the capital share factor constructed by \cite{lettau2019capital} to avoid measurement error problems. 

The innovations of capital share growth and variability are plotted in Figure \ref{fig:capitalsharesq}, and the descriptive statistics of the capital share factor and the capital share variability factor are reported in Table \ref{tab:kssq_des} in the Appendix.

\subsection{Portfolio Returns}
In our paper, the capital share factor and the variability are tested on different groups of portfolio returns. The portfolio groups we test include 25 size/BM, 10 long-term reverse (REV), 25 size/INV, and 25 size/OP sorted portfolio returns. The descriptive statistics of benchmark portfolio returns are reported in Appendix. For the multiplicative GARCH estimation, this paper takes cross-sectional averages of size/BM, REV, size/INV, and size/OP sorted portfolio returns respectively to mimic different market portfolios. All 
portfolio data are monthly data from the Kenneth R. French Data Library. The time span is January 1964 to August 2018.

%% file: chapters/Results/Results.tex
\section{Empirical Results}
\input{chapters/Results/Lambda.tex}
\subsection{Conditional Cross Sectional Regressions}
\input{chapters/Results/tvpbeta.tex}
\input{chapters/Results/BTVBSV.tex}
\input{chapters/Results/garch}

%% file: chapters/Results/Lambda.tex
\subsection{Unconditional Cross Sectional Regressions}
\begin{sidewaysfigure}
    \centering
    \includegraphics[width=\textwidth]{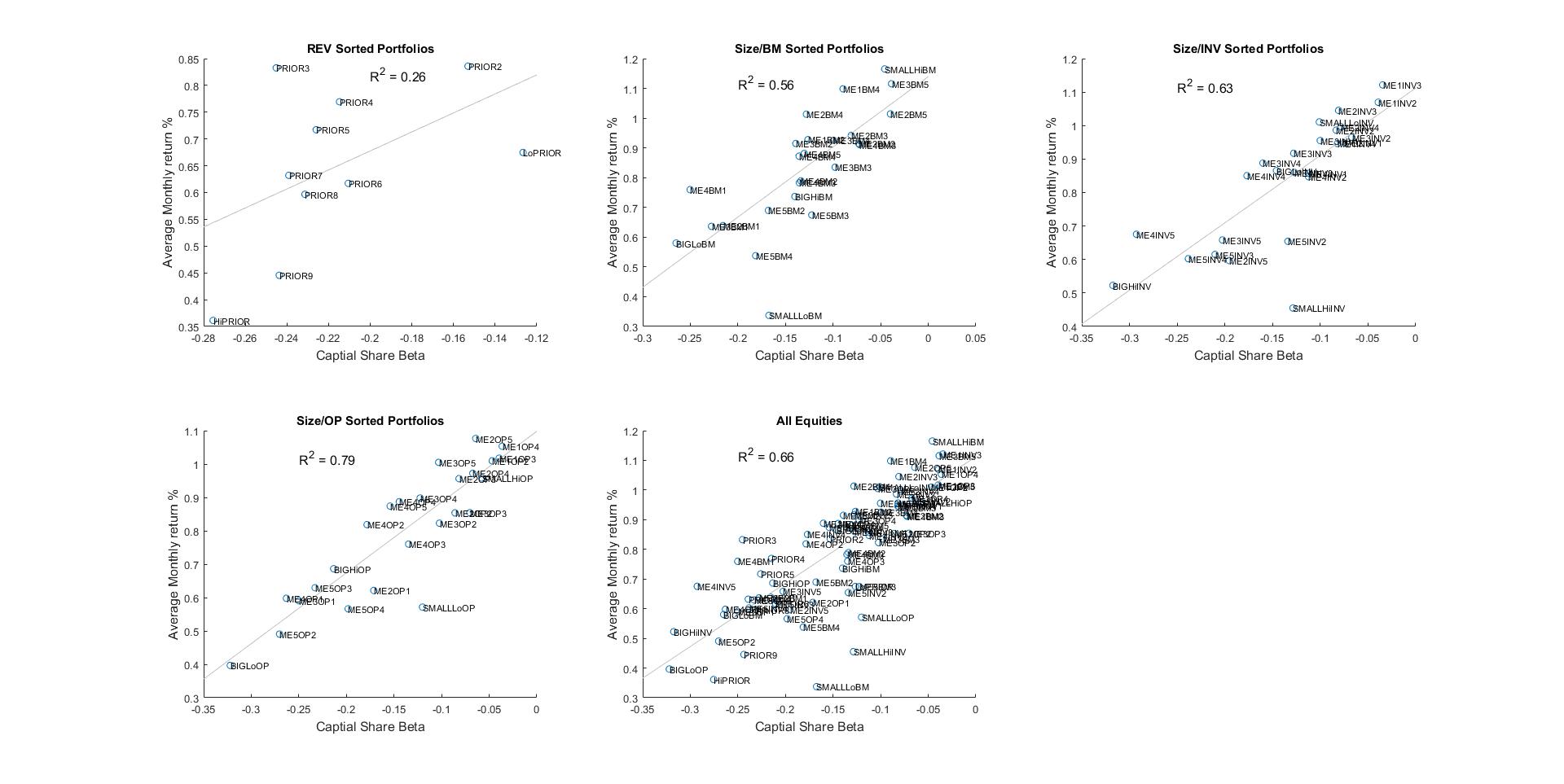}
    \begin{minipage}{.85\linewidth}{
\footnotesize \smallbreak    \caption{\textbf{Capital Share Betas.} This plot depicts the betas constructed by the F-MB regression of average portfolio returns on capital share beta. The portfolios estimated include REV, size/BM, size/INV and size/OP sorted portfolios or using all equities together. $R^2$ estimates of each regression are reported in the graph. The sample spans the period January 1974  to August 2018.}
    \label{fig:static_beta}}\end{minipage}
\end{sidewaysfigure}

To focus on testing the pricing performance of the capital share factor, our paper estimates a parsimonious capital share factor model which only contains a constant and the capital share factor. A preview of equity portfolios is shown in Figure \ref{fig:static_beta}, which plots the monthly average returns on the y-axis and the portfolio capital share betas on the x-axis. Due to the higher variation in monthly data, the $R^2$ estimates are generally lower for each portfolio class compared to the quarterly data estimates by \cite{lettau2019capital}. In addition, the $R^2$ estimated by REV sorted portfolios is 0.26 in Figure \ref{fig:static_beta}. All other $R^2$ estimated from monthly data deviate modestly from their quarterly counterparts. According to the distribution of points in the scatter plots of Figure \ref{fig:static_beta}, the model fit is high and the capital share factor has substantial explanatory power for expected returns. However, the regression lines for the portfolios deviate from 1, which indicates a potential presence of heteroskedasticity or non-linearity.

\begin{table}[ht]
\centering
\caption{\textbf{Expected Return Capital Share Risk Prices}}
\renewcommand{\arraystretch}{1}
\begin{tabular}{@{}lcccc@{}}
\\\toprule
                    & Size/BM   & REV           & Size/INV            & Size/OP            \\ \midrule                  
$\beta_0$        &  1.213**  & 1.256**   &1.170 **  & 1.189**              \\\arraybackslash 
                    &[1.068, 1.362] &[0.769, 1.731] &[1.055, 1.288]  &[1.085, 1.291]  \\
$F_{KS}$            & 2.405**  &2.560**  &  2.01**  & 2.124**              \\
                   &[1.755, 3.073]&[0.756, 4.262]  &[1.517, 2.554]   & [1.858, 2.708]\\
$\bar{R}^2$        & 0.697 &0.511  &0.721  & 0.832            \\
                    &[0.372, 0.898]&[0.000, 0.898]  &[0.429, 0.903]  & [0.618, 0.944]\\
\bottomrule
\end{tabular}
\\
\begin{tablenotes}
\item \centering
\begin{minipage}{.8\linewidth}{
\footnotesize \smallbreak
\textit{Notes: }This table reports F-MB bootstrap estimations of risk prices (\%) of the capital share factor.  The stochastic discount factor in equation (\ref{eq:ks_apx}) is tested by the single factor model stated in this table: $\beta_0$ is the constant and $F_{KS}$ is the capital share factor constructed as 12-month capital share growth. Portfolio returns used for estimation are REV, size/BM, size/INV, and size/OP sorted portfolios. Bootstrapped 95\% confidence intervals are reported in square brackets. ** denotes the estimate is significant at 5\% level. * denotes the estimate is significant at 10\% level. The sample spans the period January 1974 to August 2018.}
\end{minipage}
\end{tablenotes}
\label{tab:static_lambdas_ks}
\end{table}
The F-MB bootstrap we use is identical to that in \cite{lettau2019capital}. We therefore carry out 10000 simulations for the bootstrap process. Table \ref{tab:static_lambdas_ks} reports the risk prices estimated by the capital share factor model. In this table, all of the lower bootstrap interval bounds are above zero for capital share risk prices ($F_{KS}$), indicating the risk price estimates are all statistically significant at least at the 5\% level. For the bootstrap interval of $R^2$ estimates, the lower bound of $R^2$ for REV portfolios is 0.000, while for other portfolios are all above 0.300. Therefore, for REV portfolios, the low $R^2$ in panel B explains the insignificant capital share premium in panel A: instead of low level correlation, the high variation of correlation between portfolio returns and capital share diminishes the pricing power.

To conclude, results derived by the parsimonious unconditional capital share factor model using monthly data are consistent with the results derived by \cite{lettau2019capital} using quarterly data. The capital share risk prices are significant and positive for all equity characteristic portfolios, indicating that the capital share factor has strong pricing power. Due to different return dynamics from quarterly data, monthly returns generate lower or insignificant $\bar{R}^2$ estimates for all equity portfolios. Therefore, the cross-sectional results of the capital share risk price might vary over time dramatically, and the the increase in the frequency of the data also increases the probabilities of outliers and the variance of risk price estimates. However, the diminishing pricing power of the capital share factor in higher frequency data also indicates that this factor might be correlated with the volatility of equity returns or a potential nonlinearities in the equity return DGP.

%% file: chapters/Results/tvpbeta.tex
\subsubsection{Rolling-window Fama-MacBeth Regression}
\begin{table}[ht]

\centering
\caption{\textbf{Expected Return Capital Share Beta Rolling Regressions}}
\renewcommand{\arraystretch}{1}
\begin{tabular}{@{}lcccc@{}}
\\\toprule
                    & Size/BM& REV    & Size/INV & Size/OP            \\ \midrule
                    $\beta_0$&0.775**&0.703**&0.838**&0.813**\\
                    & (0.000)&(0.000)&(0.000)&(0.000)\\
$F_{KS}$            &-0.190*&-0.166&-0.093&-0.017\\
                    & (0.058)& (0.191)&(0.926)&(0.867) \\
$R^2$               &0.278&0.676& 0.381&0.370\\ \bottomrule
\end{tabular}
\\

\begin{tablenotes}
\item \centering
\begin{minipage}{.8\linewidth}{
\footnotesize \smallbreak
\textit{Notes:} This table reports risk prices (\%) of the capital share factor. Conditional equity premium in equation (\ref{eq:tvequitypremium}) is tested by including capital share factor $F_{KS}$, which is the 12-month capital share growth, in the mean equation. In this table, statically insignificant $F_{KS}$ rules out the possibility that the capital share factor is priced under conditional expectations. Portfolio returns used for estimation are REV, size/BM, size/INV, and size/OP sorted portfolios. The model estimated is a single capital share factor model, where $\beta_0$ is the constant and $F_{KS,t}$ is the capital share factor. the P-values are reported in parentheses below estimates. ** and * denote significance at the 5\% and 10\% levels, respectively. Sample spans the period January 1974 to August 2018.}
\end{minipage}
\end{tablenotes}

\label{tab:rw_lambdas_ks}
\end{table}

In this section, we return to estimates of the factor loadings using a rolling-window regression in the first step of the F-MB procedure. Risk prices are estimated in the same manner as the static F-MB but within each window. Table \ref{tab:rw_lambdas_ks} reports the rolling-window estimates of the parsimonious capital share factor model. As shown in this table, the capital share risk prices are insignificant for 
most equity portfolios, and all signs of risk prices are negative. The negative and insignificant risk prices show that the cross-sectional results in the second step deviate dramatically from static results (positive and significant) when estimating the first step using shorter regression windows.

\begin{sidewaysfigure}
    \centering
    \includegraphics[width=\textwidth]{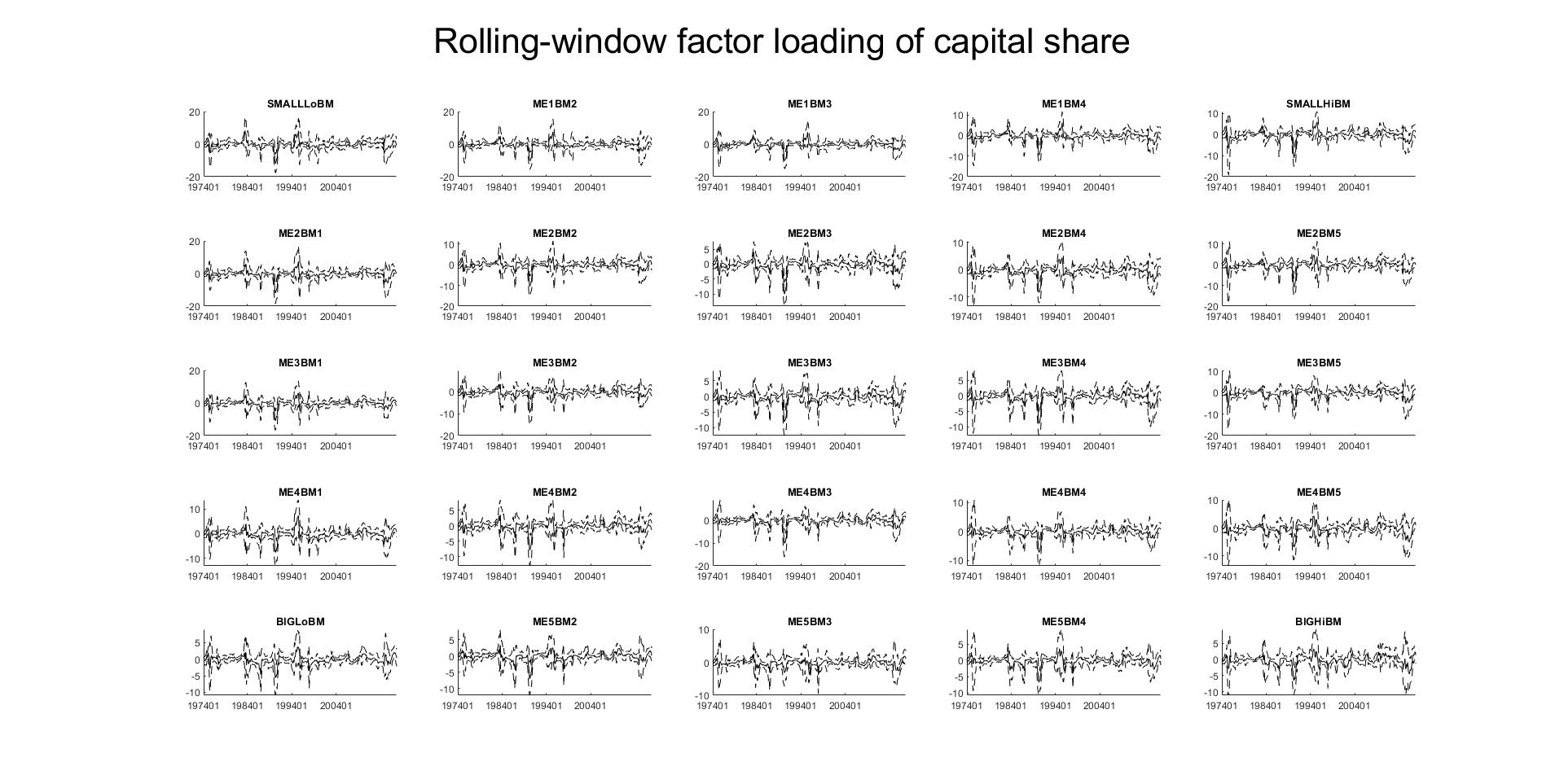}
    \begin{minipage}{.85\linewidth}{
\footnotesize \smallbreak    \caption{\textbf{12-month rolling-window estimation of capital share factor loadings, single factor model}. The factor loadings are estimated using monthly size/BM sorted portfolio returns and 12-month window length. The 95\% confidence intervals are plotted using dashed line. Sample spans January 1974 to August 2018.}
    \label{fig:rw_beta}}\end{minipage}
\end{sidewaysfigure}

Figure \ref{fig:rw_beta} plots the 12-month rolling-window estimated factor loadings of the parsimonious capital share model, and the portfolio returns estimated are size/BM sorted portfolios. As the figure shows, the factor loadings have small jumps in levels but big structural breaks in volatilities under conditional estimation. The overall level of factor loadings is centered at zero. Figure \ref{fig:rw_beta} also shows a strong volatility clustering pattern in factor loadings, which further enhances the possibility that the SDF in equation (\ref{eq:ks_apx}) and the factor model estimated might be misspecified, in the sense that the capital share factor does not enter the mean equation if we account for the time evolution of risk prices.

\begin{figure}[ht]
    \centering
    \includegraphics[width=\textwidth]{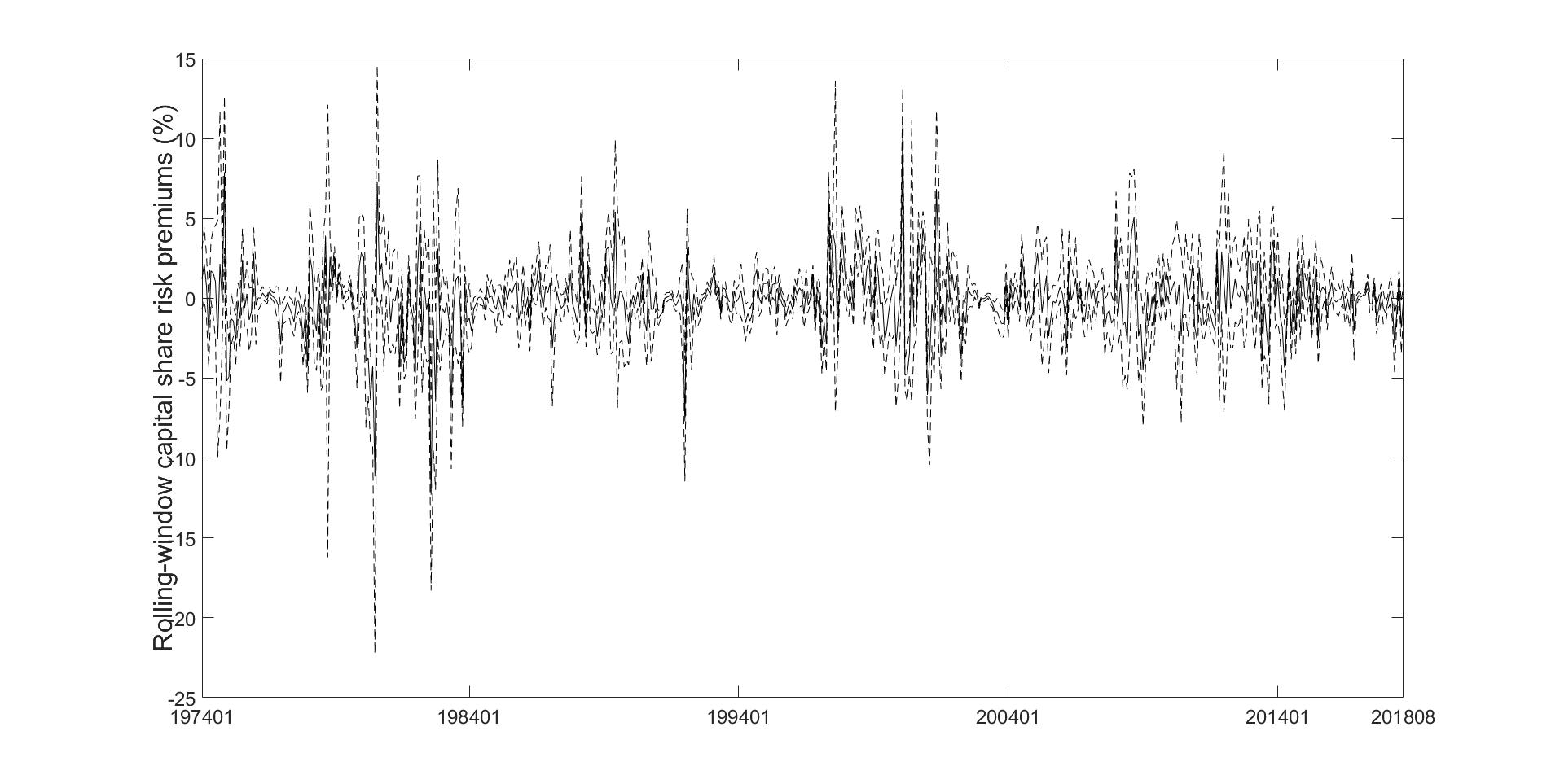}
    \begin{minipage}{.8\linewidth}{
\footnotesize \smallbreak
    \caption{\textbf{Rolling-window capital share factor risk price (\%).} Following \cite{fama1973risk} and \cite{lewellen2006conditional}, the factor loadings are estimated using monthly size/BM sorted portfolio returns and 12-month window length. The 95\% confidence intervals are plotted using dashed line. The sample spans January 1974 to August 2018.}
    \label{fig:rw_lambda}}
    \end{minipage}
\end{figure}

Figure \ref{fig:rw_lambda} plots the capital share risk prices estimated by the single factor model. This figure shows that, the time variation of risk prices is very high across the sample, and the level of risk prices witnesses frequent structural breaks. In the first step of the rolling-window F-MB estimation, the factor loadings only capture the effects caused by level changes and not the effects caused by volatility changes. In the second step estimation, the factor loadings at each time are treated as a constant, leading to a more volatile risk price series over the time when volatility varies across windows. 

Overall, the rolling-window F-MB estimates are consistent with the theoretical model in equations (\ref{eq:c_rm_inno}) and equation (\ref{eq:tvequitypremium}) in that the capital share factor loadings are centered at zero with strong volatility clustering. However, this analysis cannot rule out the potential impact of large outliers on risk price estimates due to the very short window length used. The results derived by the rolling-window F-MB procedure support accounting for structural breaks and stochastic volatility for further robustness. 

%% file: chapters/Results/BTVBSV.tex
\subsubsection{Bayesian Risk Price Estimates}
The Bayesian time-varying beta with stochastic volatility (B-TVB-SV) approach by \cite{bianchi2017macroeconomic} tackles the volatility clustering of the capital share factor loadings found by the rolling-window F-MB approach.\footnote{The prior specification of factor loading allows volatility clustering and frequent structural breaks.} the B-TVB-SV risk price estimates are more robust to outliers than those from the rolling-window F-MB estimation.

The B-TVB-SV uses 2000 burn-ins and 10000 iterations of the Markov chain Monte Carlo (MCMC) as in \cite{bianchi2017macroeconomic}, with a parsimonious capital share factor model. Following \cite{bianchi2017macroeconomic}, to robustify structural break estimates, this paper demeans all risk factors within both the training and the estimation samples to cancel out all potential bias caused by multicollinearity between the constant and the risk factors. The demeaned factors will not affect the results estimated by the B-TVB-SV since all level movements and moment conditions are retained in the sample. All Bayesian estimates passed the Geweke (\citeyear{geweke1991evaluating}) convergence diagnostic. 
\begin{sidewaysfigure}
    \centering
    \includegraphics[width=\textwidth]{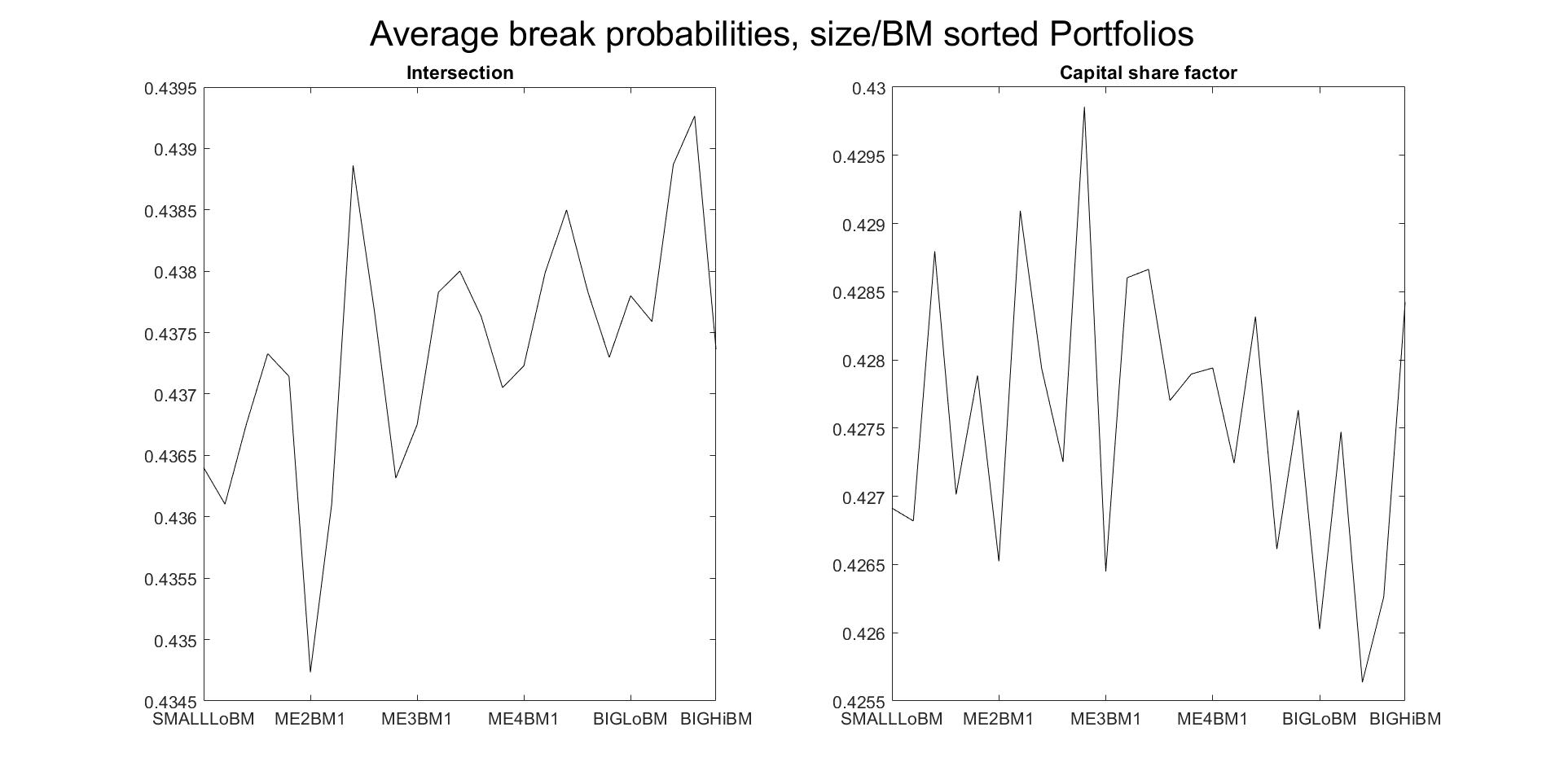}
    \begin{minipage}{.85\linewidth}{
\footnotesize \smallbreak    \caption{\textbf{B-TVB-SV average break probabilities of factor loadings, single capital share factor model.} The break probabilities are estimated using 25 size/BM sorted portfolios. Average probabilities reported are the time-average for each portfolios. The sample spans January 1964 to  August 2018. The first 10-year data in the sample is used for training, and the sample estimated covers January 1974 to  August 2018.}
    \label{fig:breaks_ks}}\end{minipage}
\end{sidewaysfigure}
 \begin{sidewaysfigure}
     \centering
     \includegraphics[width=\textwidth]{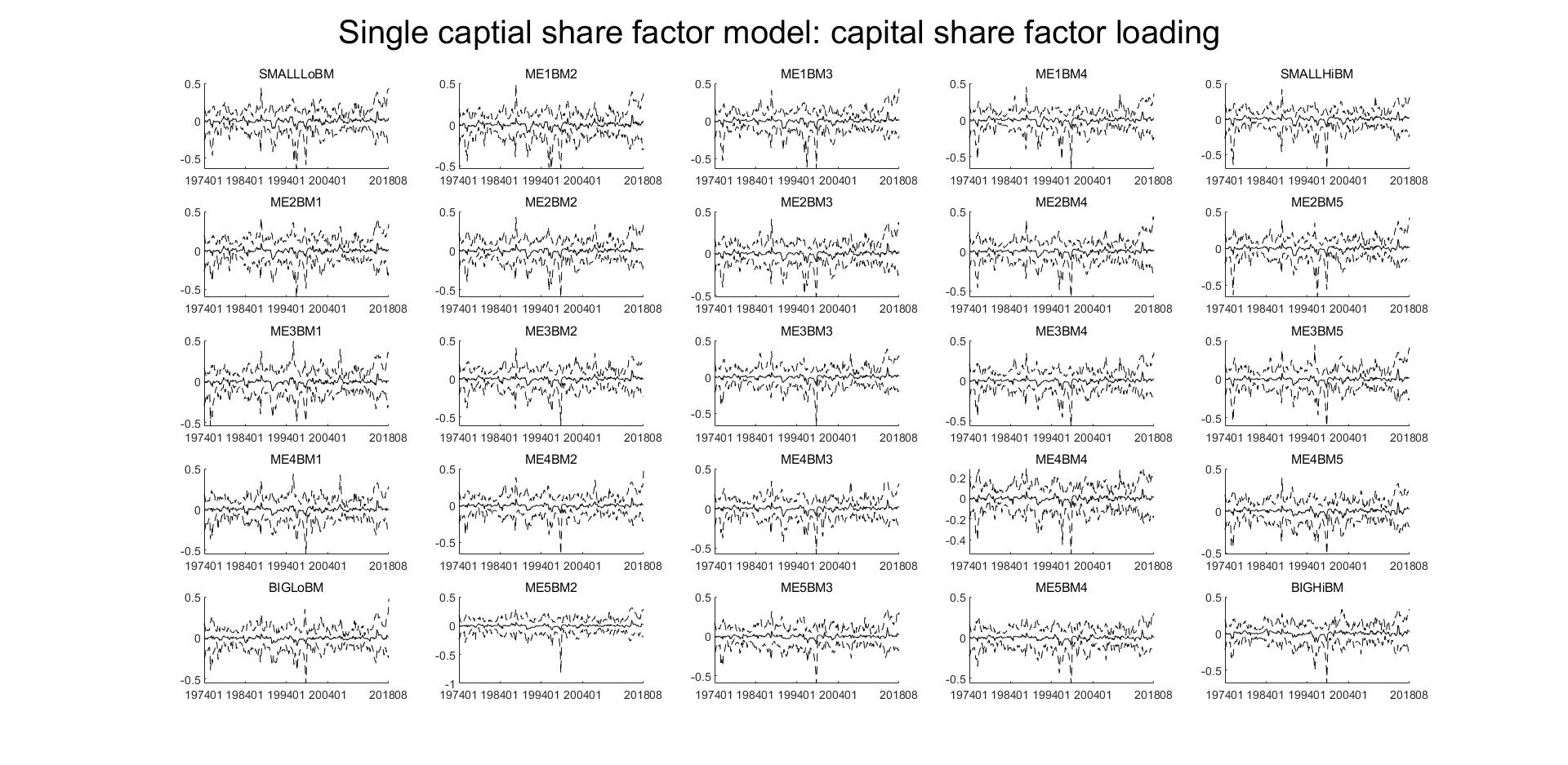}
          \begin{minipage}{.8\linewidth}{
\footnotesize \smallbreak\caption[width=\textwidth]{\textbf{B-TVB-SV capital share factor loadings.} Factor loadings are estimated by the single capital share factor model using monthly size/BM sorted portfolio returns. The 95\% confidence intervals are plotted using dashed line. The sample spans January 1964  to August 2018. The first 10-year data in the sample is used for training, and the sample estimated covers January 1974 to August 2018.}
     \label{fig:beta_single}}\end{minipage}
 \end{sidewaysfigure}
  \begin{figure}[ht]
    \centering
    \includegraphics[width=\textwidth]{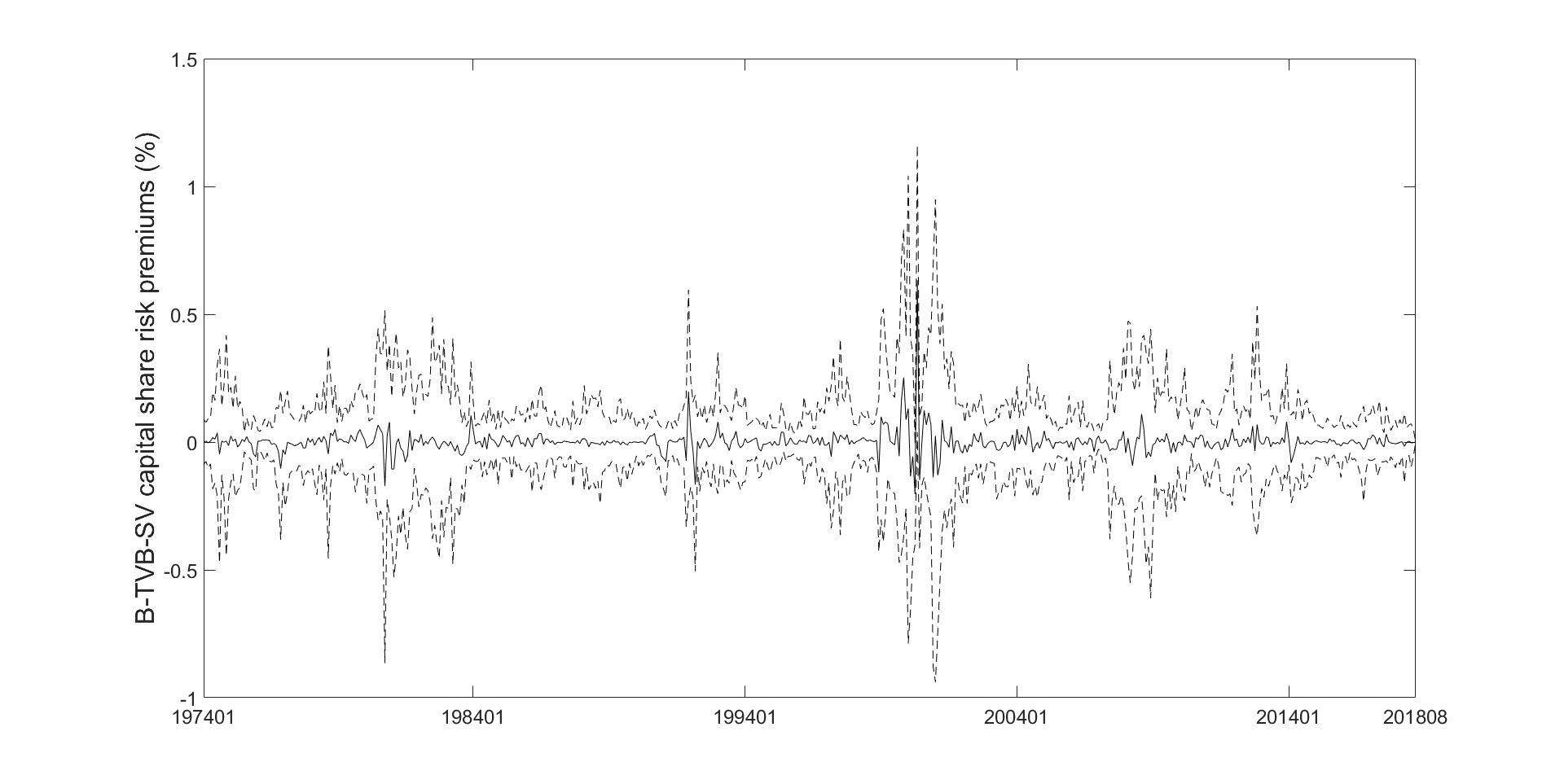}    \begin{minipage}{.85\linewidth}{
\footnotesize \smallbreak
      \caption[width=.8\textwidth]{\textbf{Bayesian capital share risk price (\%).} This figure plots risk prices estimated by the single capital share factor model using monthly size/BM sorted portfolio returns. The 95\% confidence intervals are plotted using dashed lines. The sample spans January 1964  to August 2018. The first 10-year data in the sample is used for training, and the sample estimated covers January 1974  to August 2018.}
    \label{fig:ks_lambda}}\end{minipage}
\end{figure}

As specified by the B-TVB-SV model, $\kappa_{ij,t}$ is a binary variable that equals 0 or 1. Therefore, the estimated time-average break probabilities can be viewed as a structural break test (structural breaks exist when the break probability estimates are non-zero). Figure \ref{fig:breaks_ks} plots the time-average break probabilities calculated by averaging all estimated $\kappa_{ij,t}$ in equation (\ref{eq:tvb}). To save space, this paper does not plot the time-average break probability for each portfolio. The average break probabilities of capital share are around 0.427 among the four equity portfolio classes. According to the time evolution of factor loadings stated in equation (\ref{eq:tvb}), and due to high expected value of $\kappa_{ij,t}$, $\beta_{ij,t}$ follows a jump process with frequent structural breaks over time. This finding is close to the rolling-window F-MB results reported in the previous section, and also justifies the model specification stated in equation (\ref{eq:tvb}). 

For a further robustness check of the model specification, we use plots of the capital share factor loadings in Figure \ref{fig:beta_single} and the capital factor risk price in Figure \ref{fig:ks_lambda}. From Figure \ref{fig:beta_single}, the capital share factor loadings estimated by the Bayesian method reinforces the rolling-window estimation results in that structural breaks are present both in the factor loadings and volatility. The capital share factor loadings for all portfolios are around zero. In Figure \ref{fig:ks_lambda}, the distribution of the capital share risk price is centered at zero. The mean effect of capital share is only occasionally significant, as shown by several non-zero risk price estimates.

\begin{table}[ht]
\centering
\caption{\textbf{Expect Return-Capital Share Beta Bayesian Regression}}
\renewcommand{\arraystretch}{1}
\begin{tabular}{@{}lccccccc@{}}
\\\toprule
         & Average    & Std.err    & t-stat    & p-value    & 2.5\%    & 50\%   & 97.5\%  \\ \midrule
         & \multicolumn{7}{c}{Panel A: size/BM sorted portfolios}\\
$\beta_0$ &0.832**&    0.214&    5.610& 0.000&   -9.512&    1.126&    9.403\\
$F_{KS}$ &-0.017&    0.197&    -0.085& 0.932&   -7.296&    -0.019&    7.784  \\\midrule
         & \multicolumn{7}{c}{Panel B: REV sorted portfolios}\\
$\beta_0$  &0.652**    &0.201    &3.249    &0.001   &-9.620    &0.950    &8.717\\
$F_{KS}$  &0.104    &0.242    &0.431    &0.667   &-7.932    &0.154    &8.632\\\midrule      
        & \multicolumn{7}{c}{Panel C: size/INV sorted portfolios}\\
$\beta_0$ &  0.839**    &0.215    &3.909         &0.000   &-9.404    &1.176    &9.406\\
$F_{KS}$ &-0.054    &0.157   &-0.344    &0.731   &-6.505  &-0.015    &8.066   \\\midrule      
        & \multicolumn{7}{c}{Panel D: size/OP sorted portfolio}\\
$\beta_0$ &0.801**    &0.216    &3.707 &0.000 &-9.579  &1.166 &9.302   \\
$F_{KS}$&0.085    &0.149    &0.568    &0.570   &-6.338 &0.053&8.119\\
\bottomrule
\end{tabular}\\
\begin{tablenotes}
\item \centering
\begin{minipage}{.8\linewidth}{
\footnotesize \smallbreak
\textit{Notes:} This table reports Bayesian time-varying beta with stochastic volatility proposed by \cite{bianchi2017macroeconomic}. The conditional equity premium in equation (\ref{eq:tvequitypremium}) is tested by including capital share factor $F_{KS}$, which is the 12-month capital share growth, in the mean equation. Estimates in this table are robust to time variation and volatility clustering of factor loadings. Risk prices (\%) in panels A, B, C and D  are estimated by a single capital share factor model using size/BM, REV, size/INV, and size/OP sorted portfolios, respectively. The 2.5\%,  50\%, and 97.5\% quantiles of estimated risk price distribution are included in this table. ** and * denote significance at the 5\% and 10\% levels, respectively. Data used are monthly from January 1964  to August 2018. The first 10-year data are used as training sample for hyperparameter estimation, and the sample used for estimation spans January 1974 to August 2018.}
\end{minipage}
\end{tablenotes}

     \label{tab:btvbsv_lambda_single}
\end{table}

The capital share risk prices estimated by the Bayesian method are reported in Table \ref{tab:btvbsv_lambda_single}. The risk prices in this table are estimated conditional on levels and the volatilities of both the factor loadings and the portfolio returns. In this table, the risk prices of the capital share factor are insignificant for all portfolios. With the Bayesian model specification, the factor loadings are shrunk toward zero by the weak prior when the risk factor has little effect on the level of true equity return dynamics. Therefore, capital share risk prices are insignificant when the capital share factor enters the mean of returns, even after ruling out the potential influence of outliers and stochastic volatility. Given a robust empirical evidence obtained by the Bayesian estimation, we conclude that the capital share factor does not enter the conditional stochastic discount factor of equation (\ref{eq:price_kernel}) and the mean equation of the conditional equity premium of equation (\ref{eq:tvequitypremium}).

%% file: chapters/Results/garch.tex
\subsection{A Conditional Test of the Market Return Innovation}
The capital share factor affects equity return volatility under conditional expectations, according to equations (\ref{eq:c_rm_inno}) and (\ref{eq:tvequitypremium}). The results obtained by the rolling-window F-MB and the Bayesian estimation methods rule out an impact of the capital share factor on the mean of the equity premium. We now conduct a rolling-window Multiplicative GARCH estimation as a direct test of the conditional innovation of equity return in equation (\ref{eq:c_rm_inno}). The rolling-window multiplicative GARCH specification tested in our paper is consistent with equations (\ref{eq:mfapm_kssigma}) and (\ref{eq:garch}). 
\begin{figure}[ht]
    \centering
    \includegraphics[width=\textwidth]{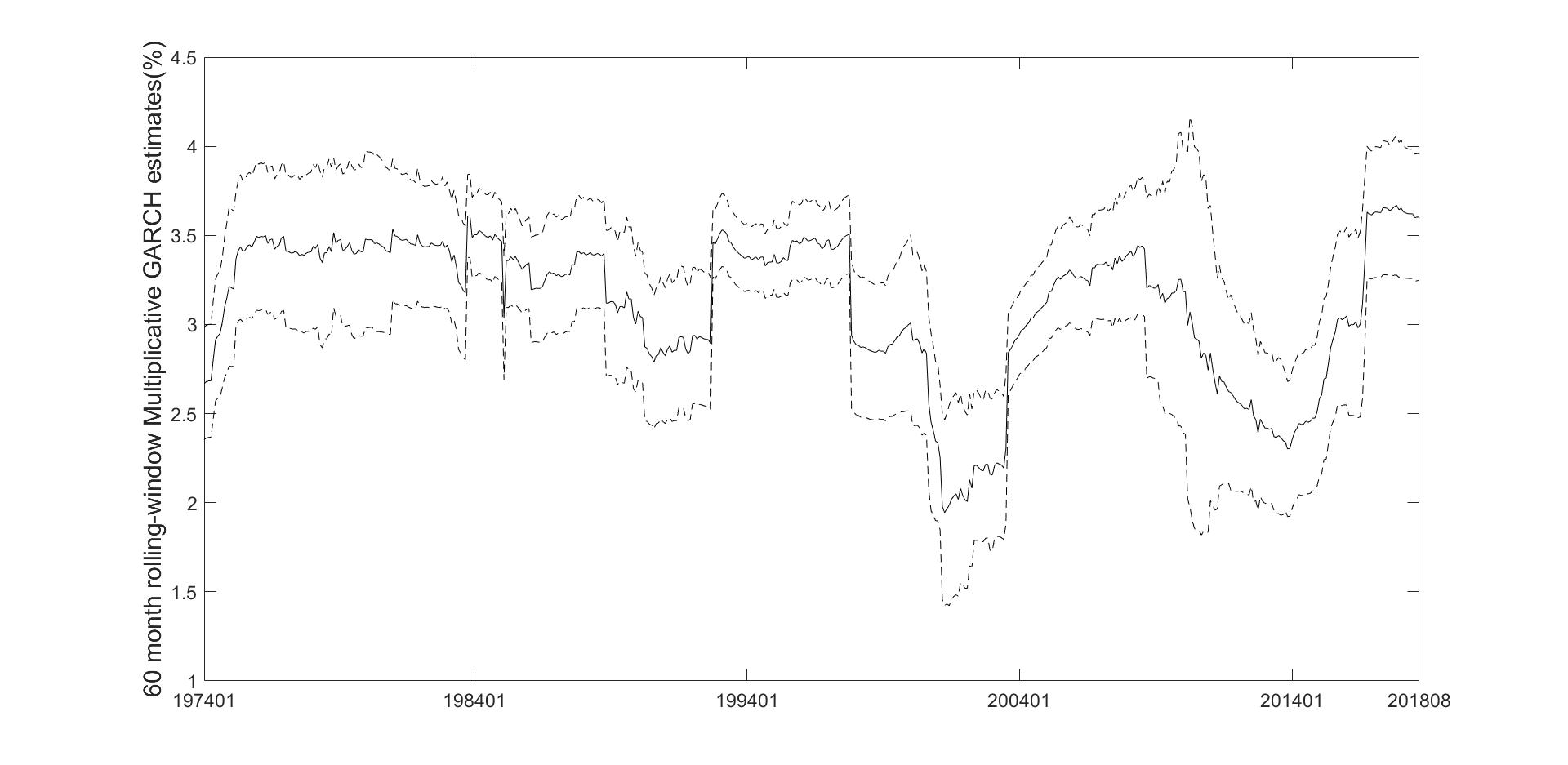}
    \begin{minipage}{.85\linewidth}{
\footnotesize \smallbreak    \caption{\textbf{60 month rolling-window multiplicative GARCH estimates (\%).} This figure shows estimates for testing conditional market return innovation in equation (\ref{eq:c_rm_inno}) in which the capital share factor enters the variance equation of equity returns. The coefficient of capital share factor is estimated using monthly average returns of size/BM sorted portfolios. The 95\% confidence intervals are plotted using dashed lines. The sample spans January 1974 to August 2018.}
    \label{fig:mgarch}}\end{minipage}
\end{figure}

The estimates for capital share are plotted in Figure \ref{fig:mgarch}. It shows that, in the variance equation, the capital share factor is always significant at the 5\% level. Compared to the Bayesian estimates reported in the previous section, the magnitude of coefficient is higher and more stable over the time horizon during which the capital share factor is insignificant in the mean equation (see Figure \ref{fig:ks_lambda}). Therefore, the capital share factor has strong a impact on the variance equation, and this variance effect dominates the mean effect under conditional estimations. This empirical evidence justifies the conditional innovation of market return in equation (\ref{eq:c_rm_inno}).

%% file: chapters/further_discussion.tex
\section{A Capital Share Variability Factor}
\begin{sidewaysfigure}
    \centering
    \includegraphics[width=\textwidth]{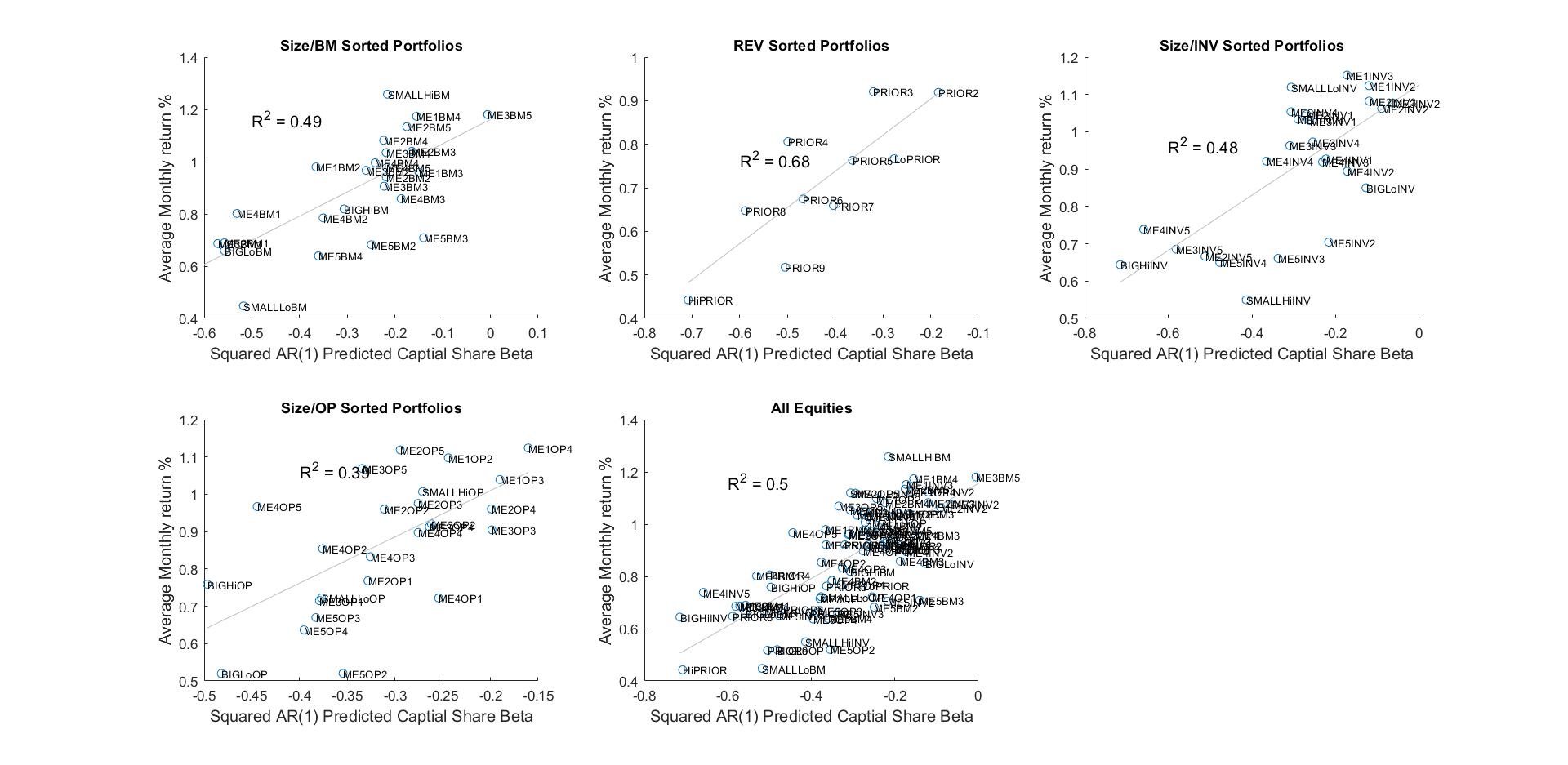}
    \begin{minipage}{.85\linewidth}{
\footnotesize \smallbreak
    \caption{\textbf{Capital share variability betas.} This plot depicts the betas constructed by the  F-MB regression of average portfolio returns on capital share variability beta. The portfolios estimated include REV, size/BM, size/INV and size/OP sorted portfolios or using all equities together. $R^2$ estimates of each regression are reported in the graph. The sample spans the period January 1974 to August 2018.}
    \label{fig:static_beta2}}\end{minipage}
\end{sidewaysfigure}

In our model, the capital share variability is a risk factor under unconditional expectations. The unconditional tests of this factor's pricing power are conducted in the same manner as earlier tests of the capital share factor. We first plot a preview of equity portfolios in Figure \ref{fig:static_beta2}. In this figure, although the average of $R^2$ is lower, the $R^2$ estimates across equity portfolios are more stable than those of Figure \ref{fig:static_beta}. Also, the slope of the regression line estimated by the capital share variability factor is closer to 1 than estimated by $F_{KS}$. Therefore, the OLS results of the capital share variability factor are robust to heteroskedasticity or nonlinearity problems.

We also estimate the capital share variability risk price using the F-MB bootstrap technique. The risk price estimates are reported in Table \ref{tab:static_lambdas_kssq}. In this table, panel A reports the single factor model that only includes a constant and capital share variability as the risk factor, and panel B reports a two factor model that includes $F_{KS}$ and the capital share variability factor for comparison. In panel A, for all equity returns, capital share variability risk prices are significant for all equity returns. The $\bar{R^2}$ estimates are stable across different portfolios and, overall, are higher than those estimated by the single capital share factor model in Table \ref{tab:static_lambdas_ks}. Note that for REV sorted portfolios the $\bar{R^2}$ estimate is insignificant in Table \ref{tab:static_lambdas_ks}, while all $\bar{R^2}$ estimates are significant in Table \ref{tab:static_lambdas_kssq}. In panel B, the $\bar{R^2}$ estimates are of similar magnitude as those in panel A. Therefore, multicolinearity might be present in the two factor model. In panel B, $F_{KS}$ is strongly dominated by the capital share variability factor. Following the inclusion of the capital share variability factor, the magnitude of capital share risk price decreases for all portfolios and becomes insignificant for REV sorted portfolios. The magnitude of capital share variability risk price also decreases following the inclusion of the capital share factor due to the colinearity between the capital share factor and its high factor volatility. However, as shown by significant capital share variability risk prices in panel B, the partial effect of the capital share variability factor remains significant in the two factor model. According to the risk price estimates in Table \ref{tab:static_lambdas_ks} and Table \ref{tab:static_lambdas_kssq}, we conclude that, under unconditional estimation, capital share variability is a stronger risk factor than capital share growth.
\begin{table}[ht]
\centering
\caption{\textbf{Capital Share Variability as a Risk Factor}}
\renewcommand{\arraystretch}{1}
\begin{tabular}{@{}lcccc@{}}
\\\toprule & Size/BM   & REV           & Size/INV            & Size/OP            \\ \midrule
& \multicolumn{4}{c}{Panel A: capital share variability}       \\ 
$\alpha$            &1.139**&1.054**&1.092**&1.181** \\
                    &[0.992, 1.280]&[0.735, 1.411]&[0.952, 1.227]&[0.959, 1.409]\\
$E(F^2_{KS})$       &8.488**&7.611**&6.966**&9.230**\\
                    &[6.277, 10.730]&[3.462, 11.79]&[4.943, 9.081]&[6.109, 12.460]\\
$\bar{R}^2$         &0.705&0.623&0.659&0.612\\
                    &[0.425, 0.888]&[0.083, 0.935]&[0.366, 0.866]&[ 0.256, 0.854]\\
\midrule
& \multicolumn{4}{c}{Panel B: two factor model}                                        \\
$\alpha$           &1.220**&1.099**&1.170**&1.197**   \\
                   &[1.054, 1.384]&[0.657, 1.544]&[1.021, 1.315]&[1.066, 1.327] \\
$F_{KS}$           &1.769**&0.980&1.768**&2.237**  \\
                   &[0.969, 2.540]&[-0.615, 2.562]&[1.001, 2.539]&[1.787, 2.707]\\
$E(F^2_{KS})$      &6.811** & 6.475**&4.423**&4.464**         \\
                   &[4.707, 8.967]&[2.093, 10.83]&[2.125, 6.787]&[2.786, 6.203]\\
$\bar{R}^2$        &0.777&0.561&0.752&0.849  \\
                   &[0.498, 0.930]&[0.000, 0.913]&[0.468, 0.922]&[0.641, 0.952]\\
\bottomrule
\end{tabular}\\
\begin{tablenotes}
\item \centering
\begin{minipage}{.8\linewidth}{
\footnotesize \smallbreak
 \textit{Notes: }This table reports F-MB bootstrap estimations of risk prices (\%) of capital share variability. Capital share variability is an unconditional risk factor according to the unconditional equity premium in equation (\ref{eq:tvequitypremium2}). In this table, portfolio returns used for estimation are REV, size/BM, size/INV, and size/OP sorted portfolios. Bootstrapped 95\% confidence intervals are reported in square brackets. ** denotes the estimate is significant at 5\% level. * denotes the estimate is significant at 10\% level. Sample spans the period January 1974 to August 2018.}
 \end{minipage}
\end{tablenotes}
\label{tab:static_lambdas_kssq}
\label{tab:ks2}
\end{table}

In summary, the factor betas in Figure \ref{fig:static_beta2} and the risk price estimates in Table \ref{tab:static_lambdas_kssq}  empirically justify the data dynamics of an unconditional equity premium (see equation (\ref{eq:tvequitypremium2})) that is positively correlated with the capital share variability factor.

%% file: chapters/conclusion.tex
\section{Conclusion}
Inspired by the work of \cite{lettau2019capital} in which U.S. asset prices are explained by capital share risks in an unconditional expected return-risk factor regression, we further investigate the role of capital share risks theoretically. Our paper develops a theoretical model of capital share risks and proposes capital share variability as an unconditional risk factor.

Following \cite{bansal2004risks}, our paper finds consumption growth volatility operates through capital share growth based upon the recursive preference utility framework developed by \cite{epstein1989substitution}. Under conditional expectations, capital share growth is found to affect the innovation of market returns but is absent from the mean equation of the equity premium. Under unconditional expectations, the capital share variability is a priced risk factor.

We first employ the Fama MacBeth bootstrap technique used by \cite{lettau2019capital} for unconditional estimations. The conditional estimations carried out in this paper include the rolling-window F-MB suggested by \cite{lewellen2006conditional}, the B-TVB-SV estimation proposed by \cite{bianchi2017macroeconomic} to test the impact of capital share factor on the conditional mean equation of equity returns, and a rolling-window multiplicative GARCH model to test the same impact but on the variance equation. 

The empirical evidence is in line with the theoretical model developed in Section 3 of our paper. Under unconditional estimations, capital share growth is found to explain the equity return dynamics. Under the rolling-window F-MB, a strong volatility clustering is found in the capital share factor loading. The capital share risk price is found insignificant but exhibits dramatic fluctuations. Under the B-TVB-SV estimation, high structural break probabilities justify the time variation of the capital share factor loadings, and the robust capital share risk price is found insignificant in the mean equation. Significant rolling-window multiplicative GARCH estimates explain the failure of the capital share factor in the mean equation of conditional equity returns: the capital share factor shows a strong multiplicative heteroskedasticity impact on the variance equation of equity return dynamics. Accordingly, we propose a capital share variability factor and test this new factor unconditionally using the F-MB bootstrap technique. We find this factor dominates the capital share factor. Therefore, the impact of the capital share factor on return volatility is the main source of its conditional pricing power.  Under unconditional expectations, capital share variability is a strong unconditional risk factor that captures the long-run movements in market volatility.

%% file: chapters/appendix/appendix.tex
\setcounter{table}{0}
\renewcommand{\thetable}{A\arabic{table}}
\setcounter{figure}{0}
\renewcommand{\thefigure}{A\arabic{figure}}
\title{Asset Prices and Capital Share Risks: A Theoretical and Empirical Contribution-Online Appendix}
\textbf{\Huge Appendix}

This appendix is not for publication and describes the Bayesian time varying beta with stochastic volatility (B-TVB-SV) specification, the detailed theoretical induction of our model, the construction of the dataset, and data basic statistics and estimations.
\section{The B-TVB-SV model specification}
 
\cite{bianchi2017macroeconomic} assumes the structural breaks are independent both across portfolio returns and over time. Equation (\ref{eq:K}) defines the structural break probabilities:
\begin{align}
    &Pr[\kappa_{ij,t}=1]=\pi_{ij}~~~~~~~~~~~~~i=1,...,N\nonumber\\
    &Pr[\kappa_{iv,t}=1]=\pi_{iv}~~~~~~~~~~~~~j=0,...,K
    \label{eq:K}
\end{align}
The probabilities $\pi_{ij}$ and $\pi_{iv}$ are sampled using a uninformative prior to retain the robustness of estimations. The priors are assumed to follow beta distributions:
\begin{align}
    &\pi_{ij}\sim Beta(a_{ij}, b_{ij})~~~~~~~~~~~~~i=1,...,N\nonumber\\
    &\pi_{iv}\sim Beta(a_{iv}, b_{iv})~~~~~~~~~~~~~j=0,...,K
    \label{eq:pi}
\end{align}
 
The structural break estimation in \cite{bianchi2017macroeconomic} uses an efficient generation of mixing variables developed by \cite{gerlach2000efficient}. In modeling intervention in dynamic mixture models, this sampling approach allows the state matrix to be singular and, hence, estimations are allowed to depend on unknown parameters. The breaks innovations $\kappa_{ij,t}$ in equation (\ref{eq:tvb}) are assumed to be conditional on the residual variance matrix ($\Sigma$), the break probability matrix of $\sigma$ ($K_\sigma$), the simulated model parameter $\theta$, excess returns $R$, and factors $F$. 
In equation (\ref{eq:sv}), $\kappa_{iv,t}$ is assumed to follow a similar innovation process to $\kappa_{ij,t}$. The conditional variance parameters of the size of the structural breaks are assumed to follow an inverted Gamma-2 distribution, of which the shape parameter is linked to the scale parameter \citep{bianchi2017macroeconomic}.    
 
The prior of the second step risk prices is a mixture of 10 random normal distributions. Priors of these normal distributions are  proposed by \cite{omori2007stochastic}. The risk price prior is as follows:
\begin{equation}
    \lambda\sim MN(\underline{\lambda},\underline{V})
    \label{eq:lambda_prior}
\end{equation}
The prior of $\tau^2$ in equation (\ref{eq:btvbsv2}) follows a inverse Gamma-2 distribution with shape parameter $\Bar{\psi_0}$ and scale parameter $\Psi$, where
\begin{equation}
    \Psi=\Psi_0+(r-\beta\lambda)'(r-\beta\lambda)
    \label{eq:psi}
\end{equation}
The risk prices are sampled conditional on the price error matrix $r-\beta\lambda$ linking the time-series regression in equation (\ref{eq:btvbsv1}) and the second-step cross-sectional regression in equation (\ref{eq:btvbsv2}). Therefore, although the risk prices are estimated in a similar manner to the F-MB procedure within each iteration, the estimated standard deviations of risk prices are robust when a firm effect is present in portfolio returns.
\input{chapters/appendix/proofappx}
\input{chapters/appendix/interpolation.tex}
\input{chapters/appendix/descriptive.tex}
\begin{table}[ht]
\centering
\caption{\textbf{Risk Factors Descriptive Statistics (\%)}}
\begin{tabular}{@{}lcccc@{}}
\toprule
     & Mean   & Median & Std.dev. & Sharpe ratio \\ \midrule
     $F_{KS}$ &&&&\\
January 1964 - January 1974  & -0.245 & -0.502 & 2.690    & -0.091       \\
January 1974 - August 2018  & 0.435  & 0.195  & 2.336    & 0.186        \\
January 1964 - August 2018  & 0.310  & 0.074  & 2.416    & 0.129        \\ \midrule
$E(F^2_{KS})$    &&&&  \\
January 1964 - January 1974 & 0.065 &0.024 &0.085& 0.764      \\
January 1974 - August 2018  & 0.051 &0.017 & 0.083 & 0.615   \\
January 1964 - August 2018  & 0.054 & 0.018&0.084&  0.643   \\ \bottomrule
\end{tabular}

\begin{tablenotes}
\item \centering
\begin{minipage}{.8\linewidth}{
\footnotesize \smallbreak
\textit{Notes:} $F_{KS}$ denotes the capital share factor and $E(F^2_{KS})$ denotes the capital share variability factor. The training sample spans January  1964 to  January  1974. The sample used for estimation spans January 1974 to August 2018. The full sample spans January 1964 to August 2018. }
\end{minipage}
\end{tablenotes}
\label{tab:kssq_des}
\end{table}

%% file: chapters/appendix/proofappx.tex
\section{Theoretical Framework}
We derive the impact of high income shareholder excess volatility on the price-consumption ratio (see equation (\ref{eq:A_2,t}) in the main text) and the price-dividend ratio (see equation (\ref{eq:A_2,m,t}) in the main text). Also, the conditional and unconditional innovation of the pricing kernel (equations (\ref{eq:price_kernel}) and (\ref{eq:m_inno2})), the conditional and unconditional innovation of equity returns (equations (\ref{eq:c_rm_inno}) and (\ref{eq:uc_rm_inno2})). The equity premium with conditional and unconditional expectations (equations (\ref{eq:tvequitypremium}) and (\ref{eq:tvequitypremium2})) in the main text are derived in this section.

With \cite{epstein1989substitution} recursive preferences, the asset pricing restrictions for gross return $R_{i,t+1}$ satisfy
 \begin{equation}
     E_t[\delta^\theta G^{-\frac{\theta}{\psi}}_{t+1}R^{-(1-\theta)}_{a,t+1}R_{i,t+1}]=1
     \label{eq:recursive_appx}
 \end{equation}
where $\theta=(1-\gamma)/(1-\frac{1}{\psi})$. In equation (\ref{eq:recursive}), $G_{t+1}$ denotes the aggregate consumption growth rate, and $R_{a,t+1}$ denotes the gross return on an asset that generates dividends that cover the aggregate shareholder consumption. $0<\delta<1$ is the time discount factor, $\gamma\geq0$ is the risk-aversion parameter, and $\psi\geq0$ is the intertemporal elasticity of substitution (IES). 

Our system equation is:
\begin{align}
    &x_{t+1}=\rho  x_{t}+\phi_e\sigma e_{t+1}\nonumber\\
    &g_{t+1}=\mu+x_{t}+w^H(1+F_{KS,t+1})\xi_{t+1}+\sigma\eta_{t+1}\nonumber\\
    &g_{d,t+1}=\mu_d+\phi x_{t}+\phi_d\sigma_{d,t+1} u_{t+1}  \label{eq:system_gappx}\\
    & ~~~~~~~~~~e_{t+1},u_{t+1},\eta_{t+1}\sim N_{i.i.d.}(0,1)~~~~~~\xi_{t}\sim N(0,\Sigma)\nonumber
\end{align}
According to \cite{bansal2004risks}, dividend growth volatility is correlated with  consumption growth volatility. Thus, $\sigma_{d,t+1}$ is partially correlated with $F_{KS,t+1}\xi_{t+1}$.

The IMRS is 
\begin{equation}
    m_{t+1}=\theta log\delta-\frac{\theta}{\psi}g_{t+1}+(\theta-1)r_{a,t+1}
    \label{eq:imrsappx}
\end{equation}
Consumption return follows:
\begin{equation}
    r_{a,t+1}=\kappa_0+\kappa_1 z_{t+1}-z_t+g_{t+1}
    \label{eq:r_apxappx}
\end{equation}
where
\begin{equation}
   z_t=A_0+A_1x_t+A_{2,t}\xi_t
   \label{eq:zappx}
\end{equation}
Following \cite{bansal2004risks}, assuming $r_{a,t+1}=r_{i,t+1}$, IMRS in equation (\ref{eq:imrsappx}) indicates:
\begin{equation}
    log\delta-\frac{1}{\psi}g_{t+1}+ r_{a,t+1}=0
    \label{eq:eqliappx}
\end{equation}
Substituting equations (\ref{eq:system_gappx}), (\ref{eq:r_apxappx}) and (\ref{eq:zappx}) into equation (\ref{eq:eqliappx}), we get:
\begin{align}
    & log\delta+(1-\frac{1}{\psi})(\mu+x_t +w^H(1+E_t(F_{KS,t+1}))E_t(\xi_{t+1})+\sigma\eta_{t+1})\nonumber\\
    &+ \kappa_0+\kappa_1 (A_0+A_1\rho  x_{t}+A_1\phi_e\sigma+E_t(A_{2,t+1})E_t(\xi_{t+1}))-(A_0+A_1x_t+A_{2,t}\xi_{t})=0
    \label{eq:46}
\end{align}
To ensure equation (\ref{eq:46}) holds, the following must hold:
\begin{align}
    &(1-\frac{1}{\psi})x_{t}+\kappa_1\rho A_1x_t-A_1x_t=0\label{eq:A1appx}\\
    &(1-\frac{1}{\psi})w^H E_t(F_{KS,t+1})E_t(\xi_{t+1})+\kappa_1E_t(A_{2,t+1})E_t(\xi_{t+1})-A_{2,t}\xi_t=0\label{eq:A2appx}
\end{align}
Notice that although the long-run expectation of $\xi_{t+1}$ is zero, this term is relatively stable between $t$ and $t+1$. Our model assumes the existence of an $r_\xi$ such that $\xi_t-r_\xi<\xi_{t+1}<\xi_t+r_\xi$ due to smoothed consumption of each income group. $r_\xi$ is a very small number which allows $\xi_{t+1}$ to deviate from $\xi_t$ while ruling out explosive growth. Therefore, $E_t(\xi_{t+1})\approx\xi_t$. Our model also assumes $E_t(A_{2,t+1})= A_{2,t}$ due to the following relationship derived from equation (\ref{eq:A2appx}):
\begin{equation}
    E_t(A_{2,t+1})=(1-\frac{1}{\psi})w^H \frac{\rho^{KS}F_{KS,t}}{\kappa_1}+A_{2,t}
\end{equation}
Assume that the value of $A_{2,t}$ equals $A_{2,0}$ at $t=0$, it is easy to solve that:
\begin{equation}
    A_{2,t}=(1-\frac{1}{\psi})w^H[\frac{\kappa_1}{\rho^{KS}(\kappa_1-\rho^{KS})}(F_{KS,0}-\frac{F_{KS,t}}{\kappa^t_1})+\frac{A_{2,0}}{\kappa^t_1}]
    \label{eq:A2tappx}
\end{equation}
According to \cite{bansal2004risks} and \cite{campbell1988dividend}, the magnitude of $\kappa_1$ is very close to 1. The value of $A_{2,t}$ is bounded by definition, thus the true $\kappa_1$ and $A_{2,0}$ are not concerns. As shown by equation (\ref{eq:A2tappx}), 
\begin{align}
    E_t(A_{2,t+1})&=(1-\frac{1}{\psi})w^H[\frac{\kappa_1}{\rho^{KS}(\kappa_1-\rho^{KS})}(F_{KS,0}-\rho^{KS}\frac{F_{KS,t}}{\kappa^{t+1}_1})+\frac{A_{2,0}}{\kappa^{t+1}_1}]\nonumber\\
    &\approx (1-\frac{1}{\psi})w^H[\frac{\kappa_1}{\rho^{KS}(\kappa_1-\rho^{KS})}(F_{KS,0}-\frac{F_{KS,t}}{\kappa^{t}_1})+\frac{A_{2,0}}{\kappa^{t}_1}]\nonumber\\
    &=A_{2,t} 
\end{align}
when $\kappa_1\approx\rho^{KS}\approx1$. Therefore, assuming $E_t(A_{2,t+1})= A_{2,t}=E(A_{2,t+1})$ is reasonable. 

Following \cite{lettau2019capital}, our paper assumes that the capital share growth rate follows an AR(1) process\footnote{The constant is not significant due to the AR(1) estimation. The magnitude of $\rho^{KS}$ is 0.947.}:
\begin{equation}
    F_{KS,t+1}=\rho_{KS}F_{KS,t}+e^{KS}_t
    \label{eq:ksinnovappx}
\end{equation}
The functional form of $A_1$ and $A_{2,t}$ can be solved:
\begin{align}
    &A_1=\frac{1-\frac{1}{\psi}}{1-\kappa_1\rho}\\
    &A_{2,t}=\frac{1-\frac{1}{\psi}}{1-\kappa_1}w^H \rho_{KS}F_{KS,t}
\end{align}
Following the same steps used in deriving the consumption premium, our paper further derives the equity premium. Equity returns have the following functional form:
\begin{equation}
    r_{m,t+1}=\kappa_{0,m}+\kappa_{1,m} z_{t+1}-z_t+g_{d,t+1}
    \label{eq:rm_apxappx}
\end{equation}
where 
\begin{equation}
    z_t=A_{0,m}+A_{1,m}x_{t}+A_{2,m,t}\xi_t
    \label{eq:rm_zappx}
\end{equation}

To further derive the equity premium $r_{m,t}$, our paper invokes the Euler condition $E[exp(m_{t+1}+r_{m,t+1})]=1$.
The following condition holds: 
\begin{align}
    \theta log\delta-\frac{\theta}{\psi}g_{t+1}+(\theta-1)r_{a,t+1}+r_{m,t+1}=0
    \label{eq:constr_rmappx}
\end{align}
To solve $A_{1,m}$ and $A_{2,m,t}$, substitute equations (\ref{eq:imrsappx}), (\ref{eq:r_apxappx}), (\ref{eq:rm_apxappx}) and (\ref{eq:rm_zappx}) into equation (\ref{eq:constr_rmappx}), collecting all terms containing $x_t$ and $\xi_t$ respectively:
\begin{align}
    &(\theta-1-\frac{\theta}{\psi})x_t+(\theta-1)(\kappa_{1}\rho -1)A_{1}x_t+\kappa_{1,m}A_{1,m}\rho x_t-A_{1,m}x_t\phi x_{t}\nonumber\\
    &=-\frac{1}{\psi}x_t+\kappa_{1,m}A_{1,m}\rho x_t-A_{1,m}x_t+\phi x_{t}=0\\
    &(\theta-1-\frac{\theta}{\psi})w^H(1+E_t(F_{KS,t+1}))E_t(\xi_{t+1})+(\theta-1)(\kappa_{1}E_t(A_{2,t+1})E_t(\xi_{t+1}) -A_{2,t}\xi_t)\nonumber\\
    &+\kappa_{1,m}E_t(A_{2,m,t+1})E_t(\xi_{t+1}) -A_{2,m,t}\xi_t\nonumber\\
    &=(\theta-1-\frac{\theta}{\psi}-\frac{1}{\psi}\rho_{KS}F_{KS,t})w^H +\kappa_{1,m}A_{2,m,t}-A_{2,m,t}=0
\end{align}
The functional form of $A_{1,m}$ and $A_{2,m,t}$ can now be solved as:
\begin{align}
    &A_{1,m}=\frac{\phi-\frac{1}{\psi}}{1-\kappa_1\rho}\\
    &A_{2,m,t}=\frac{\theta-1-\frac{\theta}{\psi}}{1-\kappa_{1,m}}w^H -\frac{w^H\rho_{KS}}{\psi(1-\kappa_{1,m})}F_{KS,t}
\end{align}
\subsubsection{Conditional on information set at time t}
The conditional innovation of consumption return is:
\begin{align}
     r_{a,t+1}-E_t(r_{a,t+1})= &\sigma\eta_{t+1}+\kappa_1A_1\phi_e\sigma e_{t+1}+[w^H(1+F_{KS,t+1})+\kappa_1A_{2,t+1}]\xi_{t+1}\nonumber\\
     &-E_t[w^H(1+F_{KS,t+1})+\kappa_1A_{2,t+1}]\xi_{t+1}\nonumber\\
     =&\sigma\eta_{t+1}+\lambda_{r,e} \sigma e_{t+1}+\lambda_{r,\xi,t+1} \xi_{t+1}\label{eq:r_inno_appx}
\end{align}
The conditional innovation of the pricing kernel is:
\begin{align}
    m_{t+1}-E_t(m_{t+1})=&(\theta-1-\frac{\theta}{\psi})\sigma \eta_{t+1}+(\theta-1)(\kappa_1A_1\phi_e) \sigma e_{t+1}\nonumber\\
    &+(\theta-1)[\kappa_1(A_{2,t+1}-E_t(A_{2,t+1}))]\xi_{t+1}\nonumber\\
    =&\lambda_{\eta} \sigma \eta_{t+1}+\lambda_{e}  \sigma e_{t+1}+\lambda_{\xi,t+1} \xi_{t+1}\label{eq:m_inno_appx}
\end{align}
In equations (\ref{eq:r_inno_appx}) and (\ref{eq:m_inno_appx}), the parameters are as follows:
\begin{align}
    \lambda_{r,e}=&~\kappa_1\frac{1-\frac{1}{\psi}}{1-\kappa_1\rho}\phi_e\\
    \lambda_{r,\xi,t+1}=&~ (w^H+\kappa_1\frac{1-\frac{1}{\psi}}{1-\kappa_1}w^H \rho_{KS})e^{KS}_{t+1}\\
    \lambda_{\eta}=& ~\theta-1-\frac{\theta}{\psi}\\
    \lambda_{e}=&~(\theta-1)(\kappa_1\frac{1-\frac{1}{\psi}}{1-\kappa_1\rho}\phi_e)\\
    \lambda_{\xi,t+1}=&~(\theta-1)(\kappa_1\frac{1-\frac{1}{\psi}}{1-\kappa_1}w^H \rho_{KS})e^{KS}_{t+1}
\end{align}
The conditional consumption premium in the presence of time-varying economic uncertainty is
\begin{align}
    E_t(r_{a,t+1}-r_{f,t})=&cov_t((m_{t+1}-E_t(m_{t+1}))(r_{a,t+1}-E_t(r_{a,t+1}))+0.5Var_t(r_{a,t+1})\nonumber\\
   = &-(\lambda_{\eta}+\lambda_{r,e}\lambda_{e}-0.5\lambda^2_{r,e}-0.5)\sigma^2\nonumber\\&+E_t(\lambda_{r,\xi,t+1}\lambda_{\xi,t+1}-0.5\lambda^2_{\xi,t+1})
    \label{eq:tvequitypremiumappx}
    \end{align}
The conditional innovation of equity return is:
\begin{align}
     r_{m,t+1}-E_t(r_{m,t+1})= &\phi_d\sigma_{d,t+1} u_{t+1}+\kappa_{1,m}A_{1,m}\phi_e\sigma e_{t+1}+\kappa_{1,m}A_{2,m,t+1}\xi_{t+1}\nonumber\\
     =&\phi_d\sigma_{d,t+1} u_{t+1}+\lambda_{m,e} \sigma e_{t+1}+\lambda_{m,\xi,t+1} \xi_{t+1}\label{eq:rm_inno_appx}
\end{align}
In equation (\ref{eq:rm_inno_appx}), the parameters are as follows:
\begin{align}
    \lambda_{m,e}=&\kappa_{1,m}\frac{\phi-\frac{1}{\psi}}{1-\kappa_1\rho}\\
    \lambda_{m,\xi,t+1}=&\kappa_{1,m}\frac{w^H\rho_{KS}}{\psi(1-\kappa_{1,m})}e^{KS}_{t+1}
\end{align}
The conditional equity premium in the presence of time-varying economic uncertainty is
\begin{align}
    E_t(r_{m,t+1}-r_{f,t})=&cov_t((m_{t+1}-E_t(m_{t+1}))(r_{m,t+1}-E(r_{m,t+1}))+0.5Var(r_{m,t+1})\nonumber\\
    =&-(\lambda_{m,e}\lambda_{e}-0.5\lambda^2_{m,e})\sigma^2+0.5\phi^2_d\sigma^2_{d,t+1}\nonumber\\&+E_t(\lambda_{m,\xi,t+1}\lambda_{\xi,t+1}-0.5\lambda^2_{m,\xi,t+1})
    \label{eq:tvequitypremium2appx}
    \end{align}
where $E_t(\lambda_{m,\xi,t+1}\lambda_{\xi,t+1}-0.5\lambda^2_{m,\xi,t+1})=0$ due to $E_t(e^{KS}_{t+1})=0$; $\sigma^2_g$ is close to $\sigma^2$ due to very small $\xi^2$. Therefore, the expected equity premium can be viewed as a constant when the model only contains capital share growth as the independent variable. The deviation of equity returns is correlated with $\sigma_{d,t+1}$ which is a function of $F_{KS,t+1}$ and $\xi_{r+1}$. In our conditional model, $F_{KS,t+1}$ is a variable that enters the variance equation.
\subsubsection{Unconditional case}
Under unconditional expectations, $E(\xi_t)=0$. Therefore, the unconditional innovation of consumption return is:
\begin{align}
     r_{a,t+1}-E(r_{a,t+1})= &\sigma\eta_{t+1}+\kappa_1A_1\phi_e\sigma e_{t+1}+[w^H(1+F_{KS,t+1})+\kappa_1A_{2,t+1}]\xi_{t+1}\nonumber\\
     =&\sigma\eta_{t+1}+\lambda_{r,e} \sigma e_{t+1}+\lambda^u_{r,\xi,t+1} \xi_{t+1}\label{eq:r_inno_appx2}
\end{align}
The unconditional innovation of the pricing kernel is:
\begin{align}
    m_{t+1}-E(m_{t+1})=&(\theta-1-\frac{\theta}{\psi})\sigma \eta_{t+1}+(\theta-1)(\kappa_1A_1\phi_e) \sigma e_{t+1}+(\theta-1)(\kappa_1A_{2,t+1})\xi_{t+1}\nonumber\\
    =&\lambda_{\eta} \sigma \eta_{t+1}+\lambda_{e}  \sigma e_{t+1}+\lambda^u_{\xi,t+1} \xi_{t+1}\label{eq:m_inno_appx2}
\end{align}
The unconditional consumption premium in the presence of time-varying economic uncertainty is
\begin{align}
    E_t(r_{a,t+1}-r_{f,t})=&cov((m_{t+1}-E_t(m_{t+1}))(r_{a,t+1}-E(r_{a,t+1}))+0.5Var(r_{a,t+1})\nonumber\\
   = &-(\lambda_{\eta}+\lambda_{r,e}\lambda_{e}-0.5\lambda^2_{r,e}-0.5)\sigma^2\nonumber\\&+E[\lambda^u_{r,\xi,t+1}\lambda^u_{\xi,t+1}-0.5(\lambda^u_{r,\xi,t+1})^2]
    \label{eq:tvequitypremiumappx2}
    \end{align}
In equations (\ref{eq:r_inno_appx2}), (\ref{eq:m_inno_appx2}) and (\ref{eq:tvequitypremiumappx2}), the parameters are as follows:
\begin{align}
    \lambda_{r,e}=&~\kappa_1\frac{1-\frac{1}{\psi}}{1-\kappa_1\rho}\phi_e\\
    \lambda^u_{r,\xi,t+1}=&~ w^H(1+F_{KS,t+1})+\kappa_1\frac{1-\frac{1}{\psi}}{1-\kappa_1}w^H \rho_{KS}F_{KS,t+1}\\
    \lambda_{\eta}=& ~\theta-1-\frac{\theta}{\psi}\\
    \lambda_{e}=&~(\theta-1)(\kappa_1\frac{1-\frac{1}{\psi}}{1-\kappa_1\rho}\phi_e)\\
    \lambda^u_{\xi,t+1}=&~(\theta-1)(\kappa_1\frac{1-\frac{1}{\psi}}{1-\kappa_1}w^H \rho_{KS}F_{KS,t+1})
\end{align}

The unconditional innovation of equity returns is:
\begin{align}
     r_{m,t+1}-E(r_{m,t+1})= &\phi_d E(\sigma_{d,t+1}) u_{t+1}+\kappa_{1,m}A_{1,m}\phi_e\sigma e_{t+1}+\kappa_{1,m}A_{2,m,t+1}\xi_{t+1}\nonumber\\
     =&\phi_d\sigma u_{t+1}+\lambda_{m,e} \sigma e_{t+1}+\lambda^u_{m,\xi,t+1} \xi_{t+1}\label{eq:rm_inno_appx2}
\end{align}
The unconditional expectation of $E(\sigma_{d,t+1})$ equals to $\sigma$ due to $E(\xi_t)=0$.
In equations (\ref{eq:rm_inno_appx2}) and (\ref{eq:tvequitypremium2appx2}), the parameters are as follows:
\begin{align}
    \lambda^u_{m,e}=&\kappa_{1,m}\frac{\phi-\frac{1}{\psi}}{1-\kappa_1\rho}\\
    \lambda^u_{m,\xi,t+1}=&\kappa_{1,m}[\frac{\theta-1-\frac{\theta}{\psi}}{1-\kappa_{1,m}}w^H -\frac{w^H\rho_{KS}}{\psi(1-\kappa_{1,m})}F_{KS,t+1}]
\end{align}
Therefore, the unconditional equity premium in the presence of time-varying economic uncertainty is
\begin{align}
    E(r_{m,t+1}-r_{f,t})=&cov((m_{t+1}-E_t(m_{t+1}))(r_{m,t+1}-E(r_{m,t+1}))+0.5Var(r_{m,t+1})\nonumber\\
    =&-(\lambda_{m,e}\lambda_{e}-0.5\lambda^2_{m,e}-0.5\phi^2_d)\sigma^2\nonumber\\&+E[\lambda^u_{m,\xi,t+1}\lambda^u_{\xi,t+1}-0.5(\lambda^u_{m,\xi,t+1})^2]
    \label{eq:tvequitypremium2appx2}
    \end{align}
where $E[\lambda_{m,\xi,t+1}\lambda_{\xi,t+1}-0.5(\lambda^u_{m,\xi,t+1})^2]$ is a function of $E(F^2_{KS})$. Given the DGP of capital share growth in equation (\ref{eq:ksinnovappx}), the $E(F^2_{KS})$ is a predicted value derived by an AR(1) model. In our unconditional model,  $E(F^2_{KS})$ is a risk factor that enters the mean equation.

%% file: chapters/appendix/interpolation.tex
\subsection{Factor Interpolation}
\noindent
This paper estimates the risk exposure and risk premium of the capital share factor in a monthly setting. However, the highest frequency of capital share data is quarterly. We interpolate capital share into monthly data due to the following reasons: 1) to avoid likely information loss when converting monthly portfolio returns into quarterly data; 2) to maintain a high degree of freedom in the training set in Bayesian estimations; 3) to avoid projection errors: in the projection process of the capital share factor, the quarterly horizon is more sensitive than the monthly horizon in terms of model missimplification \citep{lamont2001economic}. To convert the factor into monthly frequency, this paper adopts the Chow-Lin interpolation approach, which is a linear regression based model with autocorrelation in the error term     \citep{RePEc:tpr:restat:v:53:y:1971:i:4:p:372-75}.

\subsubsection{Indicator calculation}
\noindent 
The commonly used Chow-Lin interpolation \citep{RePEc:tpr:restat:v:53:y:1971:i:4:p:372-75} and other alternative interpolation approaches (see \citet{fernandez1981methodological}, \citet{litterman1983random}, etc.) are all based upon the assumption that the monthly observations of interest satisfy a multiple regression relationship with some related series. Accordingly, regression based interpolation methods require related series as indicators to capture the latent monthly movement out of a quarterly time series.
 
The capital share at time $t$, denoted by $KS_t$, can be calculated as 
\begin{equation}
    KS_t=1-LS_t
\end{equation}
under the assumption that all risk sharing across workers and stockholders is imperfect \citep{lettau2019capital}. $LS_t$ denotes labour share at time $t$.
\begin{table}[h]
\centering
\caption{\textbf{Personal income and its disposition \citep{fred}}} 
\begin{tabular}{@{}lrrrr@{}}

\toprule
\textbf{Unit: Bil. of \$}                                                                                                    & \textbf{2011:12}   & \textbf{Percentage}                & \textbf{1972:01} & \textbf{Percentage} \\ \midrule
\textbf{Personal income}                                                                                                     & 13,572.40         & 100\%                              & 898.8           & 100\%               \\ \midrule
\textbf{Compensation of employees}                                                                                           & \textbf{8,283.50} & \textbf{61\%} & \textbf{644.5}  & \textbf{72\%}       \\
\begin{tabular}[c]{@{}l@{}}Proprietors' income with inventory\\   valuation and capital consumption adjustments\end{tabular} & 1,286.10          & 9\%                                & 80.2            & 9\%                 \\
\begin{tabular}[c]{@{}l@{}}Rental income of persons with capital\\   consumption adjustment\end{tabular}                     & 508.3             & 4\%                                & 21.1            & 2\%                 \\
Personal income receipts on assets                                                                                           & 2,049.30          & 15\%                               & 122.4           & 14\%                \\
Personal current transfer receipts                                                                                 & 2,367.10 & 17\%                     & 81     & 9\%      \\
\begin{tabular}[c]{@{}l@{}}Less: Contributions for government\\   social insurance, domestic\end{tabular}                    & 922               & 7\%                                & 50.3            & 6\%                 \\
Less: Personal current taxes                                                                                                 & 1,478.80          & 11\%                               & 97.5            & 11\%                \\
Equals: Disposable personal income                                                                                           & 12,093.60         & 89\%                               & 801.3           & 89\%                \\
Less: Personal outlays                                                                                                       & 11,153.00         & 82\%                               & 694.5           & 77\%                \\
Equals: Personal saving                                                                                                      & 940.5             & 7\%                                & 106.8           & 12\%                \\\bottomrule
\end{tabular}

\begin{tablenotes}
\item \centering
\begin{minipage}{.85\linewidth}{
\footnotesize \smallbreak
\textit{Notes:} Personal income is the income obtained from provision of labour, land, and capital used in current production and the net current transfer payments received from business and government. Percentage denotes the proportion of each element in personal income. Data selected are monthly,  and covers the period from January 1972 to December 2011.}
\end{minipage}
\end{tablenotes}
\label{tab:FRED}
\end{table}

Table \ref{tab:FRED} shows the personal income and its disposition. The personal income and the compensation of employees are selected by this paper for indicator construction. An additional assumption is made to increase the robustness of the indicator, as shown in equation (\ref{eq:constant}), which is that the share of compensation of employees is constantly proportional to the labour share. 
\begin{equation}
    ES_t=\gamma_m LS_t
    \label{eq:constant}
\end{equation}

In equation (\ref{eq:constant}), $ES_t$ denotes the compensation of employees share over personal income, and $\gamma_m$ is a constant.
\smallbreak
The intuition behind the indicator selection is simple. Labour share is calculated by labour compensation divided by national income.\footnote{Labour compensation: compensation of employees in national currency.} \cite{lettau2019capital} uses the labour share of national income in the nonfarm business sector to compute capital share. However, national income is only available quarterly. Therefore, personal income is the most appropriate proxy for monthly interpolation due to its relevantly stable relationship with national income. In Table \ref{tab:FRED}, personal income refers to the broad measure of household income, and the compensation of employees denotes the gross wages paid to employees within a certain period.\footnote{Here the period is one year.} Personal income is calculated by national income minus indirect business taxes, corporate income taxes and undistributed corporate profits, then adds transfer payments.\footnote{Personal income equals to national income minus corporate profits with inventory valuation and capital consumption adjustments, taxes on production and imports less subsidies, contributions for government social insurance, net interest and miscellaneous payments on assets, business current transfer payments (net), current surplus of government enterprises, and wage accruals less disbursements, plus personal income receipts on assets and personal current transfer receipts \citep{PI}}
\cite{gomme2004measuring} show that indirect taxes and subsidies are stable over time. Hence, when studying the movement of data, the difference  between national income and personal income can be ignored, because the difference is mainly caused by indirect tax and subsidies.

\smallbreak
The calculation method for $ES_t$ is as follows:
\begin{equation}
    ES_t=\frac{Com_t}{PI_t}
    \label{eq:PS_t}
\end{equation}
where $Com_t$ denotes the compensation of employees and $PI_t$ denotes personal income.

\smallbreak
To roughly estimate $\gamma_m$, this paper assumes that $\gamma_m$ and $\gamma_q$ share the same data generation process (DGP). Quarterly compensation share $ES_q$ and labour share $LS_q$ can be used to calculate quarterly $\gamma_q$ using the following function:
\begin{equation}
    \gamma_{q}=\frac{ES_q}{LS_q}
    \label{eq:gamma_q}
\end{equation}
\begin{table}[ht]
\centering
\caption{\textbf{Descriptive Statistics of $\gamma_q$}}
\begin{tabular}{@{}ccccccc@{}}
\toprule

\textbf{Min.} & \textbf{1st Qu.} & \textbf{Median} & \textbf{Mean} & \textbf{3rd Qu.} & \textbf{Max.} & \textbf{Std.dev} \\ \midrule
1.048         & 1.087            & 1.097           & 1.099         & 1.110            & 1.154         & 0.020            \\\bottomrule
\end{tabular}

\begin{tablenotes}
\item \centering
\begin{minipage}{.7\linewidth}{
\footnotesize \smallbreak
\textit{Notes: }$\gamma_q$ is estimated by compensation of employee share in personal income over labour share (equation \ref{eq:gamma_q}). Data is quarterly and covers the sample period 1972:Q1$-$2011:Q4. $\gamma_q$ is assumed to share the same DGP as $\gamma_m$.}
\end{minipage}
\end{tablenotes}
\label{tab:gamma_q}
\end{table}
Table \ref{tab:gamma_q} shows the descriptive statistics of $\gamma_q$ calculated using equation (\ref{eq:gamma_q}). The standard deviation of $\gamma_q$ is 0.020, and the mean and median are close to each other. The dispersion of $\gamma_q$ is low according to the descriptive statistics. Therefore, monthly $\gamma_m$ can be treated as a constant according to properties of quarterly $\gamma_q$.

\begin{figure}[ht]
    \centering
    \includegraphics[width=1\textwidth]{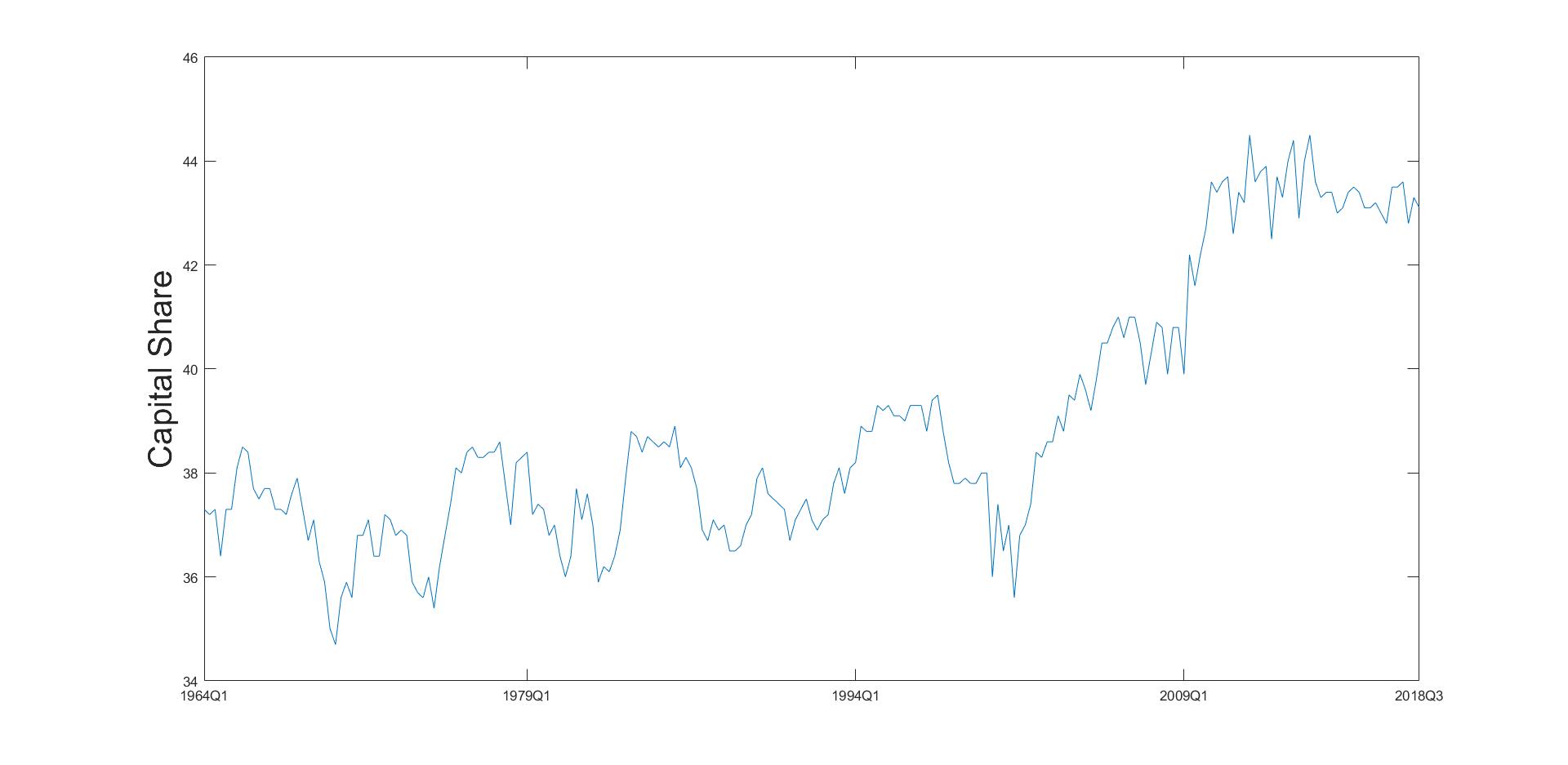}
        \begin{minipage}{.85\linewidth}{
\footnotesize \smallbreak\caption{\textbf{Capital share (quarterly).}}
        \label{fig:KS_q}}\end{minipage}
\end{figure}

\begin{figure}[ht]
    \centering
    \includegraphics[width=1\textwidth]{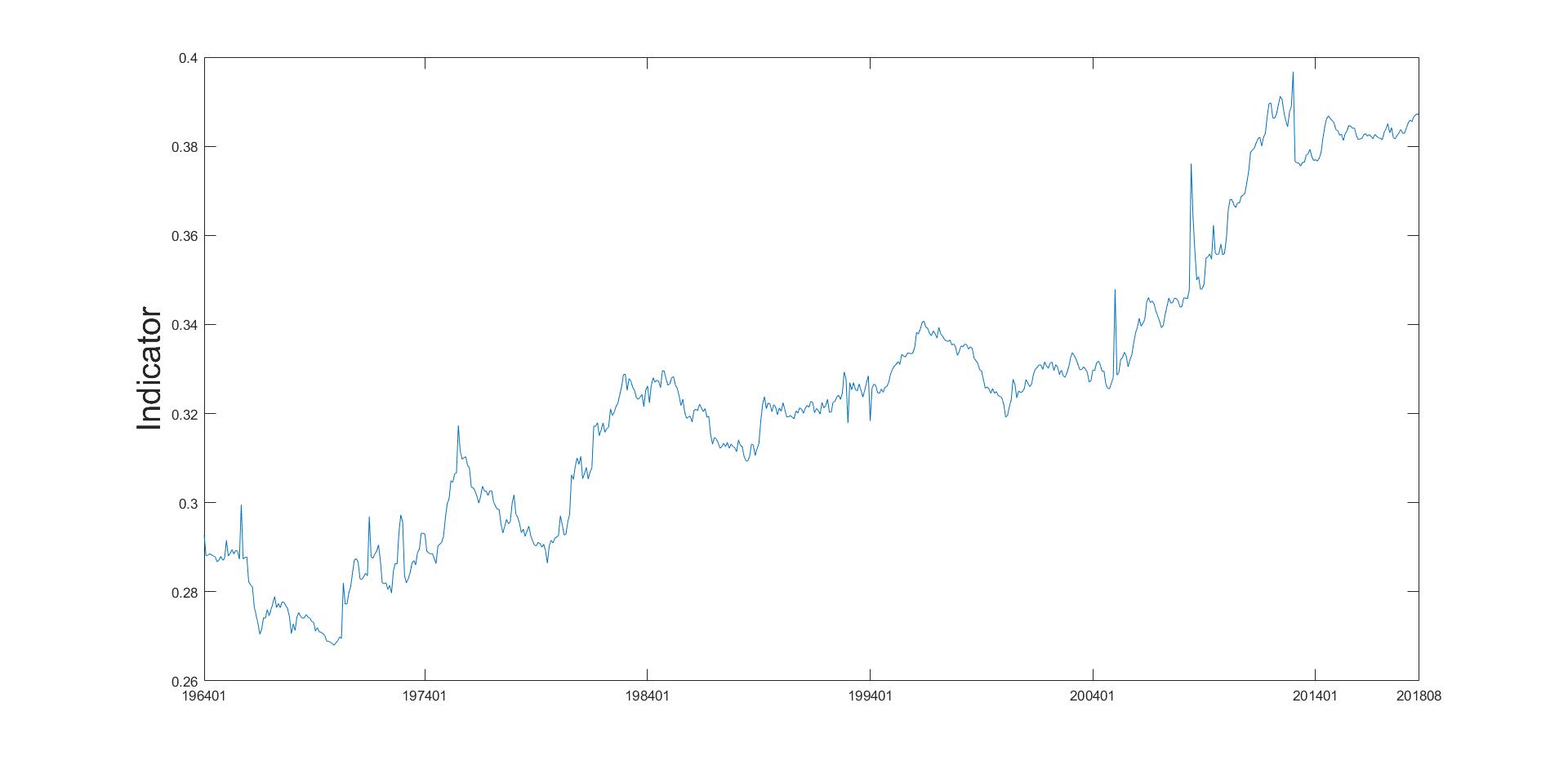}
        \begin{minipage}{.85\linewidth}{
\footnotesize \smallbreak\caption{\textbf{Indicator Dynamics}}
        \label{fig:Indicator}}\end{minipage}
\end{figure}

The movement of labour share can be represented by the share of compensation of employees. In the Chow-Lin Interpolation, the constant multiplier of the indicator is unimportant due to the regression nature of the approach. Therefore, the monthly indicator, denoted by $Ind_t$, is calculated as follows: 
\begin{equation}
    Ind_t= 1-ES_t
    \label{indicator}
\end{equation}
Figures \ref{fig:KS_q} and \ref{fig:Indicator} show the patterns of quarterly capital share factor and indicator, respectively. Although the capital share factor is overall more volatile compared to the indicator, comovements between them can still be found easily by eyeballing the two figures. 

\subsubsection{Interpolation of Capital Share}
\noindent
\cite{RePEc:tpr:restat:v:53:y:1971:i:4:p:372-75} proposes an interpolation approach based upon the assumption of a regression relationship between the latent monthly time series of interest and indicators. 
Based upon Chow-Lin method, \cite{fernandez1981methodological} and \cite{litterman1983random} approaches introduce unit roots in the error term. This paper adopts the Chow-Lin approach for interpolation and also takes potential autocorrelations in the error term of the target time series into consideration. 

Therefore, this paper assumes the following relationship holds:
\begin{equation}
    KS_{monthly}=\beta_0+\beta_{ind}Ind+\mu
    \label{chowlin}
\end{equation}
The error term $\mu$ has the following form to avoid spurious discontinuities between the last month of the previous year and the first month of the next year:
\begin{equation}
    \label{eq:mu_t}
    \mu_t=\rho~\mu_{t-1}+\epsilon_t
\end{equation}
where $KS_{monthly}$ denotes the target time series data matrix after interpolation. $Ind$ is the monthly indicator. $\mu_t$ is assumed to be an autocorrelated variable as shown in equation (\ref{eq:mu_t}). The covariance matrix of $\mu$ is denoted by $V$. $\beta_0$ and $\beta_{Ind}$ denote the constant and the coefficient of the indicator, respectively. $\rho$ is the coefficient of $\mu_{t-1}$ and captures the autocorrelation is present in the error term. $\epsilon_t$ is $i.i.d.$ and follows a standard normal distribution.
\smallbreak

The generalized least squares estimators are defined as follows in this paper:
\begin{equation}
    \label{eq:chow-lin_beta}
    \beta_{Ind}=(Ind'~V^{-1}~Ind)^{-1}~Ind'~V^{-1}~KS_{monthly}
\end{equation}
where
\begin{equation}
    \label{eq:decompoise_V}
    V=C(A'A)^{-1}C'
\end{equation}
In equation (\ref{eq:decompoise_V}), $A$ is an auxiliary matrix with the following form ($n$ equals to the quarterly data length) to factor in the autocorrelation of the error term: 
\begin{equation}
\label{matrx:A}
    A=\begin{bmatrix}(1-\rho^2)^{\frac{1}{2}}&0&0&0&\dots& & & \\
    -\rho&1&0&0&\dots& & & \\
    0&-\rho&1&0&\dots& & & \\
    \vdots&\vdots& \ddots&\ddots&&\vdots&\vdots&\vdots\\
     & & & & &\ddots&\ddots& \\
     & & & & & &-\rho&1\\
     & & & & & & &-\rho
    \end{bmatrix}_{3n\times 3n}
\end{equation}
$C$ is an $n\times 3n$ matrix with the following form:
\begin{equation}
\label{matrx:C}
    C=\begin{bmatrix}
    1 & 0  &   0    & 0 &   \dots &      &    &    &     &       \\
    0 &    0&     0&     1&     0&     0&     \dots&      &      &      \\
    &&&&&&\dots\\
          \\
   0   &  \dots  &     &       &        &      &    &       1  &   0 &    0\\
    \end{bmatrix}_{n\times 3n}
\end{equation}
\smallbreak
Grid search is used in the estimation process of the autocorrelation coefficient $\rho$. The objective function of grid searches could be the Weighted Least Square or the Log Likelihood Function. The formats of the objective functions are as follows \citep{bournay1979reflexions}:
\begin{equation}
    \label{eq:wls-obj}
    WLS= \mu' V^{-1}\mu
\end{equation}
\begin{equation}
    \label{eq:ml-obj}
    LL=-\frac{n}{2}ln(2\pi\frac{\mu' V^{-1}\mu}{n-1}) - \frac{1}{2}log(|V|) - \frac{n}{2}
\end{equation}
\smallbreak
To select proper options of the Chow-Lin interpolation, Table \ref{tab:Chow_Lin_Setting} shows the information criteria values under different settings. According to this table, the first element Chow-Lin interpolation with constant and WLS as an objective function has the lowest AIC and BIC. Hence, this paper chooses this Chow-Lin setting to generate artificial monthly capital share data.

\begin{figure}[H]
    \centering
    \includegraphics[width=1\textwidth]{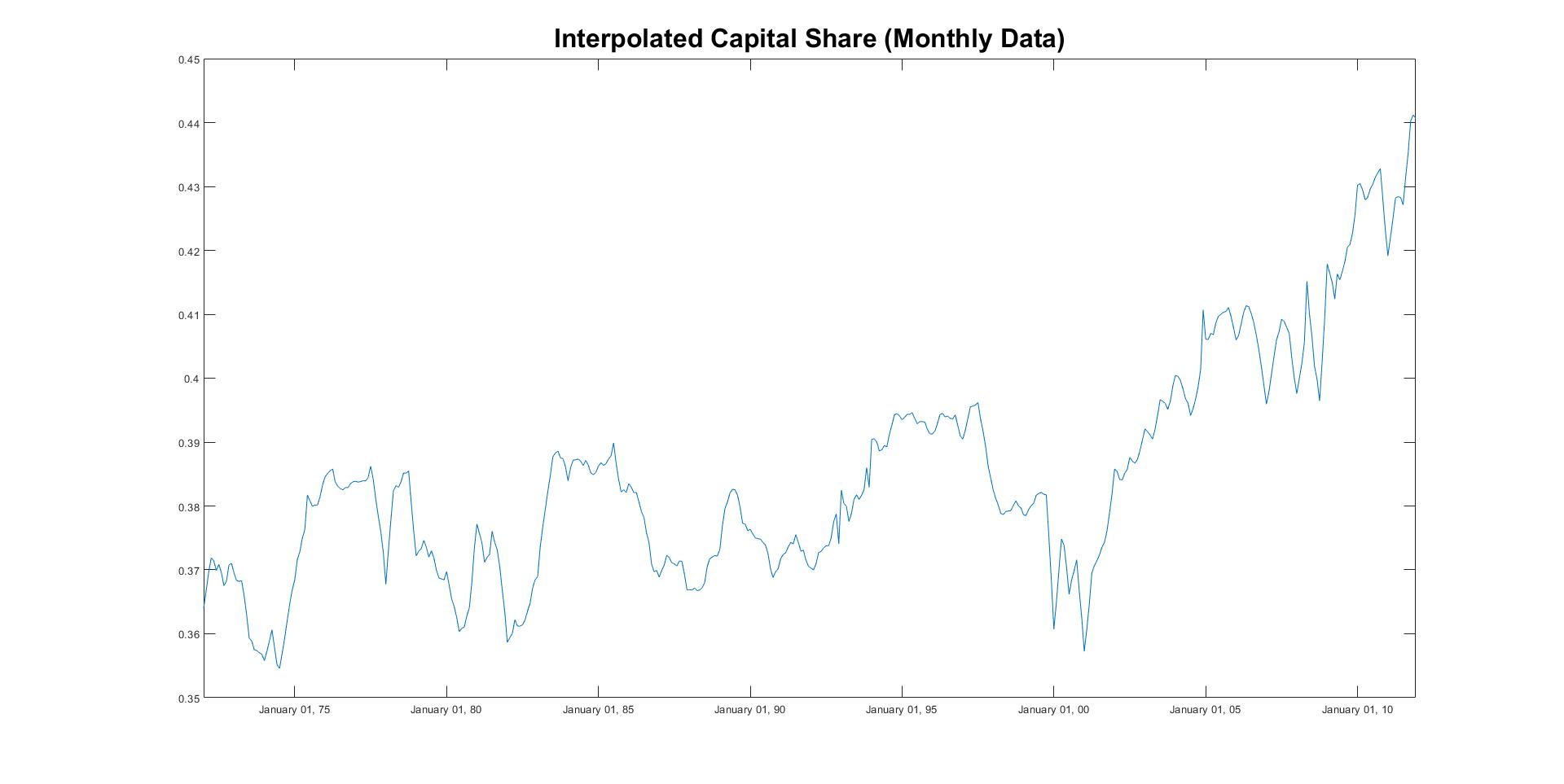}
    \caption{Interpolated Capital Share}
        \label{fig:KS_m}
\end{figure}

\begin{table}[ht]
\centering
\caption{\textbf{Information Criteria of Different Chow-Lin Settings}}
\begin{tabular}{@{}llcccc@{}}\toprule
\multicolumn{6}{c}{{Chow-Lin Settings (N=160, n=480, Quarterly to Monthly)}}                         \\ \midrule
                                   &  & \multicolumn{4}{c}{\textbf{Last Element}}                    \\ \midrule
\multirow{2}{*}{\textbf{(opc,~rl)}} &  & \multicolumn{2}{c}{\textbf{WLS}} & \multicolumn{2}{c}{\textbf{LL}} \\ \cmidrule(l){3-6} 
                                   &  & \textbf{AIC}      & \textbf{BIC} & \textbf{AIC}   & \textbf{BIC}   \\ \cmidrule(l){3-6} 
(0,~{[} {]})                        &  & -11.222           & -11.183      & -11.201        & -11.162        \\
(1,~{[} {]})                        &  & -11.384           & -11.327      & -11.349        & -11.291        \\
 \midrule
                                   &  & \multicolumn{4}{c}{\textbf{First Element}}                   \\ \midrule
\multirow{2}{*}{\textbf{(opc,~rl)}} &  & \multicolumn{2}{c}{\textbf{WLS}} & \multicolumn{2}{c}{\textbf{LL}} \\ \cmidrule(l){3-6} 
                                   &  & \textbf{AIC}      & \textbf{BIC} & \textbf{AIC}   & \textbf{BIC}   \\ \cmidrule(l){3-6} 

(0,~{[} {]})                        &  & -11.349           & -11.310      & -11.329        & -11.291        \\
(1,~{[} {]})                        &  & \textbf{-11.404}  & \textbf{-11.346}      & -11.373        & -11.315        \\\bottomrule
\end{tabular}

\begin{tablenotes}
\item \centering
\begin{minipage}{.8\linewidth}{
\footnotesize \smallbreak
\textit{Notes:} $opc$ denotes the option related to the constant. When $opc$ equals zero or one, the regression includes zero or one constant respectively. $rl$ denotes the innovational parameter. $rl =$ [~~] indicates the autocorrelation parameter $\rho$ is generated by grid search, and the calculation process adopts 100 grids of $\rho \in  [0.050, 0.999]$. WLS and LL denotes the objective function for the grid search: Weighted Least Square and Log Likelihood Function respectively.}
\end{minipage}
\end{tablenotes}

\label{tab:Chow_Lin_Setting}
\end{table}
The coefficients calculated by the Chow-Lin interpolation are shown in Table \ref{tab:chow-lin_coeffs}. The estimated constant and the indicator coefficient are both larger than two standard deviations. Although the estimated $\rho$ is close to the upper bound (0.999) of the grid search, since $\rho$ does not go beyond 1,  the conditions of partition of residuals still hold \citep{bournay1979reflexions}.
\begin{table}[ht]
\caption{\textbf{Chow-Lin coefficients under selected model specification}} 
\centering
\begin{tabular}{@{}lccc@{}}
\toprule
              & Values & Std.dev & t-statistics \\ \midrule
Constant ($\beta_0$)  & \textbf{0.192}  & 0.055   & 3.515        \\
$\beta_{Ind}$ & \textbf{0.449}  & 0.121   & 3.704        \\
$\rho$        & \textbf{0.989}  &         &              \\ 
\bottomrule
\end{tabular}

\begin{tablenotes}
\item \centering
\begin{minipage}{.8\linewidth}{
\footnotesize \smallbreak
\textit{Notes:} Bold denotes significant or feasible autocorrelation coefficients. $\beta_0$ and $\beta_{Ind}$ are both significant at 95\% confident level. The estimated autocorrelation coefficient $\rho$ is within the range of $\rho \in  [0.050, 0.999]$ for grid search, indicating no unit roots present in the error term.}
\end{minipage}
\end{tablenotes}

\label{tab:chow-lin_coeffs}
\end{table}
Figure \ref{fig:KS_m} plots the interpolated monthly capital share data.

\newpage

%% file: chapters/appendix/descriptive.tex
\section{Descriptive Statistics}
The descriptive statistics of all portfolio returns and control factors are in Tables (\ref{tab:REV_descraptive}) to (\ref{tab:sizeOP_descraptive}) below:
\begin{table}[ht]
\centering
\caption{\textbf{10 REV sorted portfolio returns (\%)}}
\begin{tabular}{@{}lcccc@{}}
\toprule
                 & \multicolumn{4}{c}{10 Size/REV sorted portfolios, value-weighted} \\ \midrule
Portfolio/Factor & Mean        & Median       & Std. dev.       & Sharpe ratio       \\ \midrule

LoPRIOR          & 1.000       & 1.135        & 7.184           & 0.139              \\
PRIOR2           & 1.152       & 1.140        & 5.674           & 0.203              \\
PRIOR3           & 1.154       & 1.375        & 5.040           & 0.229              \\
PRIOR4           & 1.039       & 1.335        & 4.656           & 0.223              \\
PRIOR5           & 0.996       & 1.165        & 4.422           & 0.225              \\
PRIOR6           & 0.907       & 1.240        & 4.270           & 0.213              \\
PRIOR7           & 0.892       & 1.105        & 4.227           & 0.211              \\
PRIOR8           & 0.881       & 1.155        & 4.375           & 0.201              \\
PRIOR9           & 0.750       & 0.855        & 4.676           & 0.161              \\
HiPRIOR          & 0.676       & 0.795        & 5.403           & 0.125            \\\bottomrule 
\end{tabular}

\begin{tablenotes}
\item \centering
\begin{minipage}{.7\linewidth}{
\footnotesize \smallbreak
\textit{Notes:} Data frequency is monthly. Time span of data is from January 1964 to August 2018.}
\end{minipage}
\end{tablenotes}
\label{tab:REV_descraptive}
\end{table}

\begin{table}[ht]
\centering
\caption{\textbf{25 Size/BM sorted portfolio returns (\%)}}
\begin{tabular}{@{}lcccc@{}}
\toprule
                 & \multicolumn{4}{c}{25 Size/BM sorted portfolios, value-weighted} \\ \midrule
Portfolio/Factor & Mean         & Median       & Std. dev.       & Sharpe ratio      \\ \midrule

SMALLLoBM        & 0.681       & 1.060       & 7.854          & 0.087            \\
ME1BM2           & 1.213       & 1.523       & 6.849          & 0.177            \\
ME1BM3           & 1.192       & 1.254       & 5.934          & 0.201            \\
ME1BM4           & 1.407      & 1.450       & 5.644          & 0.249            \\
SMALLHiBM        & 1.491      & 1.485       & 5.946          & 0.251            \\
ME2BM1           & 0.923       & 1.376       & 7.094          & 0.130            \\
ME2BM2           & 1.174       & 1.456       & 5.924          & 0.198            \\
ME2BM3           & 1.273       & 1.530       & 5.374          & 0.237            \\
ME2BM4           & 1.315       & 1.528       & 5.197          & 0.253            \\
ME2BM5           & 1.367       & 1.788       & 5.964          & 0.229            \\
ME3BM1           & 0.920       & 1.546       & 6.515          & 0.141            \\
ME3BM2           & 1.20       & 1.505       & 5.383          & 0.223            \\
ME3BM3           & 1.19       & 1.486       & 4.943          & 0.230            \\
ME3BM4           & 1.268       & 1.442       & 4.855          & 0.261            \\
ME3BM5           & 1.414       & 1.524       & 5.587          & 0.253            \\
ME4BM1           & 1.035       & 1.157       & 5.823         & 0.178            \\
ME4BM2           & 1.018      & 1.215       & 5.052          & 0.201            \\
ME4BM3           & 1.091       & 1.354       & 4.906          & 0.222            \\
ME4BM4           & 1.229       & 1.420       & 4.720          & 0.260            \\
ME4BM5           & 1.210       & 1.423       & 5.626          & 0.215            \\
BIGLoBM          & 0.893       & 0.998       & 4.569          & 0.195            \\
ME5BM2           & 0.915       & 1.073       & 4.375          & 0.209            \\
ME5BM3           & 0.942       & 1.215       & 4.231          & 0.223            \\
ME5BM4           & 0.872       & 0.995       & 4.581          & 0.190            \\
BIGHiBM          & 1.052       & 1.319       & 5.326          & 0.198      \\\bottomrule     
\end{tabular}

\begin{tablenotes}
\item \centering
\begin{minipage}{.7\linewidth}{
\footnotesize \smallbreak
\textit{Notes:} Data frequency is monthly. Time span of data is from January 1964 to August 2018.}
\end{minipage}
\end{tablenotes}
\label{tab:sizeBM_descraptive}
\end{table}
\begin{table}[ht]
\centering
\caption{\textbf{25 Size/INV sorted portfolio returns (\%)}}
\begin{tabular}{@{}lcccc@{}}
\toprule
                 & \multicolumn{4}{c}{25 Size/INV sorted portfolios, value-weighted} \\ \midrule
Portfolio/Factor & Mean         & Median       & Std. dev.       & Sharpe ratio      \\ \midrule

SMALLLoINV       & 1.353       & 1.376       & 7.183          & 0.188            \\
ME1INV2          & 1.357       & 1.413       & 5.599          & 0.242            \\
ME1INV3          & 1.385       & 1.631       & 5.603          & 0.247            \\
ME1INV4          & 1.266       & 1.561       & 5.926          & 0.214            \\
SMALLHiINV       & 0.783       & 1.028       & 7.049          & 0.111            \\
ME2INV1          & 1.280       & 1.636       & 6.298          & 0.203            \\
ME2INV2          & 1.296       & 1.586       & 5.209          & 0.249            \\
ME2INV3          & 1.315       & 1.490       & 5.197          & 0.253            \\
ME2INV4          & 1.287       & 1.595       & 5.633          & 0.228            \\
ME2INV5          & 0.900       & 1.235       & 6.893          & 0.131            \\
ME3INV1          & 1.263       & 1.460       & 5.673          & 0.223            \\
ME3INV2          & 1.310       & 1.475       & 4.775          & 0.274            \\
ME3INV3          & 1.196       & 1.383       & 4.761          & 0.251            \\
ME3INV4          & 1.206       & 1.495       & 5.273          & 0.229            \\
ME3INV5          & 0.919       & 1.307       & 6.441          & 0.143            \\
ME4INV1          & 1.160       & 1.455       & 5.318         & 0.218            \\
ME4INV2          & 1.127       & 1.388       & 4.709          & 0.239            \\
ME4INV3          & 1.152       & 1.402      & 4.620          & 0.249            \\
ME4INV4          & 1.154      & 1.269       & 4.867          & 0.237            \\
ME4INV5          & 0.972       & 1.224       & 6.240          & 0.156            \\
BIGLoINV         & 1.083       & 1.125       & 4.554          & 0.238            \\
ME5INV2          & 0.937       & 0.920       & 3.957          & 0.237            \\
ME5INV3          & 0.894       & 1.000       & 4.066          & 0.220            \\
ME5INV4          & 0.883       & 1.045       & 4.379          & 0.202            \\
BIGHiINV         & 0.877       & 1.113       & 5.390          & 0.163   \\\bottomrule        
\end{tabular}

\begin{tablenotes}
\item \centering
\begin{minipage}{.7\linewidth}{
\footnotesize \smallbreak
\textit{Notes:} Data frequency is monthly. Time span of data is from January 1964 to August 2018.}\end{minipage}
\end{tablenotes}
\label{tab:sizeINV_descraptive}
\end{table}
\begin{table}[ht]
\centering

\caption{\textbf{25 Size/OP sorted portfolio returns (\%).}}
\begin{tabular}{@{}lcccc@{}}
\toprule
                 & \multicolumn{4}{c}{25 Size/OP sorted portfolios, value-weighted} \\ \midrule
Portfolio/Factor & Mean        & Median       & Std. dev.       & Sharpe ratio      \\ \midrule

SMALLLoOP        & 0.955       & 0.980        & 7.218           & 0.132             \\
ME1OP2           & 1.331       & 1.471        & 5.791           & 0.230             \\
ME1OP3           & 1.273       & 1.581        & 5.583           & 0.228             \\
ME1OP4           & 1.357       & 1.505        & 5.739           & 0.237             \\
SMALLHiOP        & 1.240       & 1.477        & 6.546           & 0.190             \\
ME2OP1           & 1.001       & 1.522        & 6.944           & 0.144             \\
ME2OP2           & 1.193       & 1.633        & 5.640           & 0.212             \\
ME2OP3           & 1.209       & 1.510        & 5.244           & 0.230             \\
ME2OP4           & 1.194       & 1.298        & 5.509           & 0.217             \\
ME2OP5           & 1.352       & 1.698        & 6.143           & 0.220             \\
ME3OP1           & 0.948       & 1.208        & 6.535           & 0.145             \\
ME3OP2           & 1.154       & 1.485        & 5.091           & 0.227             \\
ME3OP3           & 1.138       & 1.384        & 4.866           & 0.234             \\
ME3OP4           & 1.146       & 1.286        & 5.106           & 0.225             \\
ME3OP5           & 1.302       & 1.554        & 5.753           & 0.226             \\
ME4OP1           & 0.955       & 1.077        & 6.044           & 0.158             \\
ME4OP2           & 1.087       & 1.391        & 5.057           & 0.215             \\
ME4OP3           & 1.066       & 1.250        & 4.720           & 0.226             \\
ME4OP4           & 1.131       & 1.293        & 4.833           & 0.234             \\
ME4OP5           & 1.200       & 1.558        & 5.307           & 0.226             \\
BIGLoOP          & 0.753       & 1.051        & 5.444           & 0.138             \\
ME5OP2           & 0.753       & 0.926        & 4.412           & 0.171             \\
ME5OP3           & 0.903       & 1.033        & 4.325           & 0.209             \\
ME5OP4           & 0.870       & 1.127        & 4.357           & 0.200             \\
BIGHiOP          & 0.992       & 1.106        & 4.273           & 0.232    \\\bottomrule        
\end{tabular}

\begin{tablenotes}
\item \centering
\begin{minipage}{.7\linewidth}{
\footnotesize \smallbreak
\textit{Notes:} Data frequency is monthly. Time span of data is from January 1964 to August 2018.}\end{minipage}
\end{tablenotes}
\label{tab:sizeOP_descraptive}
\end{table}